\documentclass[12pt]{article}
\pdfoutput=1
\usepackage{latexsym, graphicx,cite,mathrsfs}
\usepackage{amsmath}
\usepackage{amssymb}
\usepackage{amsthm}
\usepackage{subcaption}

\allowdisplaybreaks

\usepackage[top = 1. in,bottom = 0.93 in, left = 1 in, right=1 in]{geometry}
\usepackage[colorlinks=true, linkcolor=blue, bookmarks=true]{hyperref}
\newcommand{\arXiv}[1]{\href{http://www.arXiv.org/abs/#1}{arXiv:#1}}
\newcommand{\doi}[2]{\href{http://dx.doi.org/#2}{#1}}

\makeatletter
\renewcommand\section{\@startsection {section}{1}{\z@}%
                  {-3.5ex \@plus -1ex \@minus -.2ex}%nn
                  {2.3ex \@plus.2ex}%
                  {\normalfont\large\bfseries}}
\renewcommand\subsection{\@startsection{subsection}{2}{\z@}%
                   {-3.25ex\@plus -1ex \@minus -.2ex}%
                   {1.5ex \@plus .2ex}%
                   {\normalfont\bfseries}}
\makeatother

\numberwithin{equation}{section}

\newcommand{\beq}{\begin{equation}}
\newcommand{\eeq}{\end{equation}}
\newcommand{\ber}{\begin{array}}
\newcommand{\eer}{\end{array}}
\newcommand{\del}{\partial}

\newcommand{\de}{\delta}
\newcommand{\la}{\lambda}
\newcommand{\ka}{\kappa}
\newcommand{\bea}{\begin{eqnarray}}
\newcommand{\eea}{\end{eqnarray}}
\newcommand{\ER}{Erd\H{o}s-R\'enyi }
\renewcommand{\Re}{\operatorname{Re}}
\renewcommand{\Im}{\operatorname{Im}}
\newcommand{\Tr}{\mathrm{Tr}}
\newcommand{\bph}{\pmb{\phi}}

\begin{document}
\begin{titlepage}
\begin{flushright}
\phantom{arXiv:yymm.nnnn}
\end{flushright}
\vspace{5mm}
\begin{center}
{\LARGE\bf Random matrices with row constraints and\vspace{5mm}\\ eigenvalue distributions of graph Laplacians}\\
\vskip 17mm
{\large Pawat Akara-pipattana$^{a,b}$ and Oleg Evnin$^{c,d}$}
\vskip 7mm
{\em $^a$ UFR Physique, Facult\'e des Sciences, Universit\'e Paris Cit\'e, Paris, France}
\vskip 3mm
{\em $^b$ DISAT, Politecnico di Torino, Torino, Italy}
\vskip 3mm
{\em $^c$ Department of Physics, Faculty of Science, Chulalongkorn University,
Bangkok, Thailand}
\vskip 3mm
{\em $^d$ Theoretische Natuurkunde, Vrije Universiteit Brussel and\\
The International Solvay Institutes, Brussels, Belgium}
\vskip 7mm
{\small\noindent {\tt akarapawat@gmail.com, oleg.evnin@gmail.com}}
\vskip 20mm
\end{center}
\begin{center}
{\bf ABSTRACT}\vspace{3mm}
\end{center}

Symmetric matrices with zero row sums occur in many theoretical settings and in real-life applications. When the offdiagonal elements of such matrices are i.i.d.~random variables and the matrices are large, the eigenvalue distributions converge to a peculiar universal curve $p_{\mathrm{zrs}}(\la)$ that looks like a cross between the Wigner semicircle and a Gaussian distribution. An analytic theory for this curve, originally due to Fyodorov, can be developed using supersymmetry-based techniques.

We extend these derivations to the case of sparse matrices, including the important case of graph Laplacians for large random graphs with $N$ vertices of mean degree $c$. In the regime $1\ll c\ll N$, the eigenvalue distribution of the ordinary graph Laplacian (diffusion with a fixed transition rate per edge) tends to a shifted and scaled version of $p_{\mathrm{zrs}}(\la)$, centered at $c$ with width $\sim\sqrt{c}$. At smaller $c$, this curve receives corrections in powers of $1/\sqrt{c}$ accurately captured by our theory. For the normalized graph Laplacian (diffusion with a fixed transition rate per vertex), the large $c$ limit is a shifted and scaled Wigner semicircle, again with corrections captured by our analysis.

\vfill

\end{titlepage}

\section{Introduction}

Symmetric matrices with zero row sums (and other linear constraints on the entries of individual rows) arise naturally in a wide range of applications.
In \cite{zrs}, we find an ambitious list of various areas of research where such matrices come up,
ranging from the physics of Goldstone bosons, amorphous solids and nano-wire junctions to studies of financial markets and Google search algorithms. The most direct motivation for our current study comes from the question of evaluating eigenvalue distributions of random graph Laplacians, closely linked to zero-row-sum matrices.

When matrices with row constraints are treated in the context of random matrix theory, the first natural assumption is that the offdiagonal entries are independent identically distributed (i.i.d.)\ variables, and the diagonal entries are fixed by the constraints. It may appear somewhat surprising that, when this prescription is applied to zero-row-sum matrices, and the size of the matrices is taken to be large,
the eigenvalue distribution does {\it not} converge to the familiar Wigner semicircle. An intuitive explanation is that, for an $N\times N$ matrix, if the offdiagonal entries are i.i.d.\ with mean 0 and variance 1, the diagonal entries are fixed in terms of the offdiagonal row-sums, and are typically of magnitude $\sqrt{N}$. As a result, the diagonal entries are much bigger than the offdiagonal ones, and the diagonalization process works very differently from the GOE ensemble that produces the Wigner semicircle.

There is some analogy, mentioned in particular in \cite{zrs}, between the zero-row-sum ensemble and the Rosenzweig-Porter ensemble \cite{RP} (see \cite{Guhr,RP2022} for more contemporary treatments). In the latter, the offdiagonal entries of a real symmetric matrix and its diagonal entries are all independent random variables, but the variance scales differently with $N$ for the diagonal and offdiagonal entries. If the diagonal entries are sufficiently large, one breaks Wigner's universal behavior, and transitions, in the limit of extremely large diagonal entries, to a Gaussian eigenvalue distribution. While this analogy appears valid, it remains qualitative, since the diagonal entries in the zero-row-sum ensemble are large but not independent from the offdiagonal ones. In fact, as an aside to the main considerations of this paper, we will propose an explicit family of ensembles whose eigenvalue distributions interpolate between the Wigner semicircle and the Gaussian distribution, while the zero-row-sum ensemble appears directly at an intermediate interpolation parameter value.

A theory of eigenvalue distributions of random matrix ensembles with linear row constraints can be 
effectively developed using methods derived from statistical field theory \cite{statFT,statFTneu}.
These methods have a respectable history of applications to random matrix problems, see \cite{Guhr,RP2022,BR,MF,FM,Fyodorov,euclRM,SC,SMFL,PN,cavity,kuehn,spectrum,AC,TO,resdist} for a sampler of related literature. More specifically, we shall focus on supersymmetry-based techniques \cite{efetov,wegner} that introduce auxiliary integrals over anticommuting variables and rewrite the eigenvalue density (or more precisely, the matrix resolvent) in terms of an integral admitting a saddle point evaluation at large $N$. Fyodorov and Mirlin proposed in \cite{MF,FM} a powerful variation of this method
(see \cite{Fyodorov,TO} for further developments) that introduces functional integration into the game
and derives very effectively an integral saddle point equation for the matrix resolvent in ensembles that are not easily tractable by other methods. In \cite{Fyodorov}, Fyodorov applied this method to real symmetric matrices with zero row sums, obtaining an analytic theory for their eigenvalue distribution. Our goal in the first two sections of this article will be to give a pedagogical review of the Fyodorov-Mirlin method and Fyodorov's solution for the eigenvalue distribution of the zero-row-sum ensemble.

We remark that in the decades that have passed since \cite{Fyodorov}, rigorous mathematical results have been obtained \cite{free,dense,lessdense}
 within the {\it free probability} program
for dense random matrices in the 
zero-row-sum ensemble and related ensembles. In this context, the limiting eigenvalue distribution of large zero-row-sum real symmetric matrices
may be characterized as the free convolution of the standard Gaussian distribution and the Wigner semicircle law \cite{dense}.
These considerations have led, in particular, to a proof of the large $N$ equivalence \cite{free} of the zero-row-sum and Rosenzweig-Porter ensembles. (We mention additionally the intriguing related branch of research that deals with eigenvalue distributions of non-Hermitian matrices subject
to row constraints \cite{nonherm1,nonherm2,nonherm3}.)

The main purpose of this article is to extend the considerations for zero-row-sum ensembles to sparse matrices. This includes the important case of random graph Laplacians. A graph with $N$ vertices labelled by $i=1..N$ is commonly specified by the {\it adjacency matrix} $A_{ij}$ whose diagonal entries are 0 and whose offdiagonal entries are $1$ if a graph edge connects vertices $i$ and $j$ and 0 otherwise. We can also introduce the diagonal {\it degree matrix} $\mathbf{D}$ whose offdiagonal entries are 0 and each diagonal entry $D_{ii}\equiv \sum_j\! A_{ij}$ represents the degree of vertex $i$. An \ER random graph of mean degree $c$ (at large $N$) is obtained by treating the offdiagonal entries of $\mathbf{A}$ as i.i.d.\ random variables that equal 1 with probability $c/N$ and 0 with probability $1-c/N$.

Graph Laplacians can be defined through diffusion or random walks on graphs. First, consider a random walker that, within a small time interval $\Delta t$, traverses any edge that starts at its current position with probability $\ka\,\Delta t$. In that case, the probability $p_i(t)$ for the random walker to be found at vertex $i$ is governed by the equation
\beq
\frac{dp_i}{dt}=-\ka\sum_j L_{ij} p_j,
\eeq
where
\beq\label{ordLa}
\mathbf{L}\equiv\mathbf{D}-\mathbf{A}
\eeq
is the ordinary {\it graph Laplacian}. One could define a different random walk where, within a small time interval $\Delta t$, the random walker leaves its current location with probability $\ka\,\Delta t$, and if it does leave, it jumps to one of the nearest neighbors of the current location with equal probability. In that case, the probability $p_i(t)$ is governed, in terms of the more convenient variable $\tilde p_i\equiv \sqrt{D_{ii}} p_i$, by the equation
\beq
\frac{d\tilde p_i}{dt}=-\ka\sum_j \mathcal{L}_{ij} \tilde p_j,
\eeq
where
\beq\label{normLa}
 \pmb{\mathcal{L}}\equiv\mathbf{I}-\mathbf{D}^{-1/2}\mathbf{A}\mathbf{D}^{-1/2},
\eeq
with $\mathbf{D}^{-1/2}$ denoting the square root of the (pseudo)inverse of $\mathbf{D}$. A common name for $ \pmb{\mathcal{L}}$ is the {\it normalized graph Laplacian}.

Graph Laplacians have been discussed from a wide variety of perspectives \cite{chung,VM}, both purely mathematical and applied. A far-reaching example is seen in \cite{neuro}, where eigenvalue distributions are studied for Laplacians of graphs representing neural connectivities in the brains of small mammals.
Graph Laplacians also play a significant role in discrete geometry \cite{discr1,discr2}. Given the importance of graph Laplacians, it is somewhat surprising that few analytic studies of the distributions of their eigenvalues are seen in the literature, especially for the physically important case of large graphs of finite mean degree. One important exception is the last section of \cite{kuehn}. It focuses, however, not on recovering a closed-form analytic expression for the distribution curve, but rather on constructing an effective algorithm inspired by population dynamics that reproduces this curve as
an infinite superposition of Gaussians with fluctuating variance.

Our aim in this article is to close the existing gap in the literature and take steps toward an analytic theory
of eigenvalue distributions of graph Laplacians for large graphs of finite mean degree. After reviewing the Fyodorov-Mirlin method in the subsequent two sections, we shall turn to this problem, first for the ordinary graph Laplacian and then for the normalized graph Laplacian.

%%%%%%%%%%%%%%%%%%%%%%%%%%%%%%%%%%%%%%%

\section{The Fyodorov-Mirlin method}

%%%%%%%%%%%%

\subsection{Resolvents and supersymmetry} 

We start with spelling out the general principles for analyzing random matrix eigenvalue distributions with statistical field theory techniques, and then gradually specialize our treatment to the supersymmetry-based version of these techniques, finally introducing the Fyodorov-Mirlin method.

We shall be interested in random real symmetric $N\times N$ matrices $\mathbf{M}$ with a subset of entries given by i.i.d.\ random variables and the remaining entries expressed linearly and deterministically through those random entries. For such matrices, an effective approach to computing the density $p(\la)\equiv N^{-1}\sum_k\de(\lambda-\la_k)$ of eigenvalues $\lambda_k$ is based on expressing it through the matrix resolvent $\Tr[(\mathbf{M}-z\mathbf{I})^{-1}]$ using the Sokhotski-Plemelj formula 
$1/(x\pm i0)=P(1/x)\mp i\pi\de(x)$:
\beq\label{resolv}
p(\la)=-\frac1{\pi N}\Im\left\{\langle\Tr(\mathbf{M}-z\mathbf{I})^{-1}\rangle\right\}\Big|_{z=\la-i0}=-
\frac1{\pi}\Im\left\{\langle(\mathbf{M}-z\mathbf{I})^{-1}_{11}\rangle\right\}\Big|_{z=\la-i0},
\eeq
where the angular brackets represent averaging over the $\mathbf{M}$-ensemble, and the second equality assumes that this ensemble is symmetric under renumbering of rows and columns of $\mathbf{M}$.

An advantage of the resolvent representation (\ref{resolv}) is that the matrix inverse $(\mathbf{M}-z\mathbf{I})^{-1}$ can be effectively rewritten using Gaussian integrals so that, after such transformation has been implemented, the expression factorizes over the entries of $\mathbf{M}$, and then the $\mathbf{M}$-averaging becomes straightforward in many ensembles with statistically independent matrix entries. A simple relation to consider as a starter is
\beq\label{Gaussbsn}
(\mathbf{M}-z\mathbf{I})^{-1}_{11}\propto \sqrt{\det(\mathbf{M}-z\mathbf{I})}\int d\bph \,(i \phi_1^2)\, e^{i\sum_{kl}\phi_k M_{kl} \phi_l-iz\sum_k \phi_k^2},
\eeq
where $\bph$ is an $N$-dimensional vector with components $\phi_i$, and the proportionality sign will mean throughout this article that we are ignoring purely numerical factors (in this case, $z$-independent), which can always be recovered at the end of our computations by normalizing probability distributions to 1. Note that the integral in (\ref{Gaussbsn}) is convergent when the imaginary part of $z$ is negative, which agrees with the specification of $z$ in (\ref{resolv}).

The representation in (\ref{Gaussbsn}) is still not quite what we need because of the awkward determinant factor. There are various approaches to dealing with this factor, but the one we shall adopt throughout this article is introducing further integrals over anticommuting (Grassmannian) variables and making use of supersymmetry-based techniques \cite{efetov,wegner}. To this end, we introduce two $N$-vectors $\pmb{\xi}$ and $\pmb{\eta}$ whose components $\xi_i$ and $\eta_i$ satisfy the relations
\beq\label{Berezin}
\xi_i\xi_j=-\xi_j\xi_i,\qquad \xi_i^2=0,\qquad \int d\xi_i \xi_j=\de_{ij},\qquad \int d\xi_i=0,
\eeq
plus similar relations for $\eta_i$, while all components of $\pmb{\xi}$ also anticommute with the components of $\pmb{\eta}$. (To get the signs fixed unambiguously, one should also specify that the differentials likewise anticommute, and one needs to place each variable next to its own differential before applying the above integration rules.) The integration rules, originally due to Berezin, are a convenient formal prescription that, in particular, results in the following valuable formula valid for any $N\times N$ matrix $\pmb{\mathcal{M}}$:
\beq
\det \pmb{\mathcal{M}}=\int d\pmb{\xi}\, d\pmb{\eta}\, e^{\sum_{kl}\xi_k\mathcal{M}_{kl}\eta_l}.
\eeq
(To prove this formula, one expands the exponential as a Taylor series, and then Berezin's integration rule picks out exclusively the piece of polynomial degree $N$ in the components of $\pmb{\mathcal{M}}$, while the anticommuting nature of $\pmb{\xi}$ and $\pmb{\eta}$ antisymmetrizes this piece to make it equal to the determinant.) Armed with this relation, as well as the trivial Gaussian formula 
\beq\label{Gausschi}
1\propto \sqrt{\det(\mathbf{M}-z\mathbf{I})}\int d\pmb{\chi} \, e^{i\sum_{kl}\chi_k M_{kl} \chi_l-iz\sum_k \chi_k^2},
\eeq
where $\pmb{\chi}$ is an $N$-dimensional vector with commuting components, we can upgrade (\ref{Gaussbsn}) to a Gaussian representation for the resolvent no longer plagued by determinant factors:
\beq\label{GaussSUSY}
(\mathbf{M}-z\mathbf{I})^{-1}_{11}\propto \int d\bph\,d\pmb{\chi}\, d\pmb{\xi}\, d\pmb{\eta}\,(i\phi_1^2)\, e^{i\sum_{kl}M_{kl} (\phi_k \phi_l+\chi_k \chi_l+\xi_k\eta_l)-iz\sum_k (\phi_k^2+\chi_k^2+\xi_k\eta_k)},
\eeq
This formula\footnote{Note that it is in principle possible to evaluate the Grassmannian integrals in each formula in this article, rewriting everything completely in terms of ordinary numbers and ordinary integrals. Such representations would, however, be much more awkward and bulky than the formulas written using the Grassmannian notation. Anticommuting variables thus simply provide a very convenient shorthand for writing formulas for ordinary commuting variables, just like complex numbers provide a convenient shorthand for writing certain relations involving real numbers.} can be written more compactly by introducing the `superspace' notation which amounts to joining $\phi_k$, $\chi_k$, $\xi_k$ and $\eta_k$ into a supervector 
\beq\label{Psicomp}
\Psi_k=(\phi_k, \chi_k, \xi_k, \eta_k)^T,
\eeq
with its conjugate $\Psi_k^\dagger$ defined using the `supermetric' $S$ as
\beq
\Psi_k^\dagger\equiv \Psi_k^T\,S,\qquad S\equiv\left(\begin{matrix}\hspace{2mm}1\hspace{2mm}&0&0&0\\0&1&0&0\\0&0&0& -1/2\\0&0&1/2&0\end{matrix}\right).
\eeq
(Note that $\Psi_1^\dagger\Psi_2=\Psi_2^\dagger\Psi_1$.)
With this notation,
\beq\label{Gausssuperv}
(\mathbf{M}-z\mathbf{I})^{-1}_{11}\propto \int d\pmb{\Psi}\,(i\phi_1^2)\, e^{i\sum_{kl}M_{kl} \Psi^\dagger_k \Psi_l-iz\sum_k \Psi^\dagger_k\Psi_k}.
\eeq
At this point, we have attained complete factorization over the entries of $\mathbf{M}$ and averaging over $\mathbf{M}$ can be performed entry-by-entry. This is very convenient when working with ensembles where the entries of $\mathbf{M}$ are independent random variables. 

The result of averaging over $\mathbf{M}$ depends on the concrete ensemble, but we shall write it schematically as
\beq\label{Mav}
\langle(\mathbf{M}-z\mathbf{I})^{-1}_{11}\rangle\propto \int d\pmb{\Psi}\,(i\phi_1^2)\, e^{-\frac1{2N}\sum_{kl}C( \Psi_k, \Psi_l)-iz\sum_k \Psi^\dagger_k\Psi_k},
\eeq
where the specific form of the function $C$ should be derived for each concrete model. To see how this structure emerges, assume first that
the entries of  $\mathbf{M}$ are independent (up to the constraints imposed by the symmetries of the matrix $\mathbf{M}$, real and symmetric in our case). When we factorize
(\ref{Gausssuperv}) as $\prod_{kl}e^{i M_{kl} \Psi^\dagger_k \Psi_l}$ and perform averaging over the components of $\mathbf{M}$,  we obtain a product of factors each of which only depends on a single pair $(\Psi_k, \Psi_l)$, leading to the structure in (\ref{Mav}). When some of the entries of  
$\mathbf{M}$ are expressed through others, as in the presence of linear row constraints, the situation may potentially be more complicated, but as we shall see through direct evaluation of the average in the ensembles that interest us in this paper, the structure will always be as in (\ref{Mav}). We shall therefore focus on exploring this structure.

The integral in (\ref{Mav}) is in the form of a (super)vector model. Such models are in principle always solvable in the large $N$ limit by introducing a finite number of scalar auxiliary variables \cite{largeN}, whereupon one obtains an integral with a saddle point structure where $N$ serves as the saddle point parameter. In practice, a large number of auxiliary fields is necessary to implement this method in more complicated models, resulting in unwieldy multidimensional saddle point equations. We shall take a different route in the considerations of this paper, one that gives a more effective approach to the large $N$ limit of (\ref{Mav}).
 
%%%%%%%%%%%%

\subsection{The functional saddle point}

A problem with (\ref{Mav}) is that $C( \Psi_k, \Psi_l)$ depends on pairs of supervectors $\Psi_i$. If, instead, we had a sum of terms each of which only depended on one $\Psi_i$, the integrand would factorize over the different components of $\pmb{\Psi}$ and would be straightforwardly evaluated. The Fyodorov-Mirlin method \cite{MF} introduces a clever trick to deal with this issue (which should incidentally be more broadly appreciated in our opinion). While the construction involves a functional integral over a function from supervectors to numbers that we will call $g(\Psi)$, and this may seem like an unpleasant complication, the fact is that after this transform has been implemented, there is an explicit saddle point structure at large $N$ in the resulting functional integral, so that one obtains an explicit integral equation for the saddle point configuration $g_*(\Psi)$, and the eigenvalue distribution is in turn expressed through this saddle point configuration.

In practice, one relies on the following functional Gaussian integral
\beq\label{MFtransfrm}
\int \mathcal{D}g \exp\left[-\frac{N}2\int d\Psi\,d\Psi'\, g(\Psi) \,C^{-1}(\Psi,\Psi')\,g(\Psi')+i\sum_k g(\Psi_k)\right]=e^{-\frac1{2N}\sum_{kl}C( \Psi_k, \Psi_l)},
\eeq
where $C^{-1}$ is the inverse of $C$ in the sense of integral convolution: 
\beq
\int d\Psi\, C(\Psi_1,\Psi)\,C^{-1}(\Psi,\Psi_2)=\de(\Psi_1-\Psi_2).
\eeq
Note that the $\de$-function of a supervector $\Psi=(\phi,\chi,\xi,\eta)^T$ is
\beq
\de(\Psi)=-\xi\,\eta\,\de(\phi)\de(\chi),
\eeq
so that
\beq
\int d\Psi' F(\Psi') \,\de(\Psi-\Psi')=F(\Psi)
\eeq
for any function $F$, as follows from the integration rules (\ref{Berezin}). 

If (\ref{MFtransfrm}) is substituted into (\ref{Mav}), the integral over $\pmb{\Psi}$ factorizes into $N$ identical copies of an integral over a single component of $\pmb{\Psi}$ (the first copy is slightly modified by the insertion of $\phi_1^2$), and the expression can be restructured as follows:
\beq
\label{afterMF}
\begin{split}
&\langle(\mathbf{M}-z\mathbf{I})^{-1}_{11}\rangle\propto \int \mathcal{D}g\,e^{-N\hspace{0.1mm}S[g]}\,\frac {\int d\Psi\,(i\phi^2)\,e^{ig(\Psi)-iz\Psi^\dagger\Psi}}{\int d\Psi\,e^{ig(\Psi)-iz\Psi^\dagger\Psi}},\\
&S[g]\equiv\frac{1}2\int d\Psi\,d\Psi'\, g(\Psi) \, C^{-1}(\Psi,\Psi')\,g(\Psi')-\ln\left(\int d\Psi\,e^{ig(\Psi)-iz\Psi^\dagger\Psi}\right).
\end{split}
\eeq
There is an evident `saddle point' structure due to the presence of the large factor $N$ in the exponent, and one expects that the integral is dominated at large $N$ by the stationary points of the functional $S[g]$ defined by $\de S/\de g(\Psi)=0$, which can be written out explicitly as
\beq\label{sddlpsi}
g(\Psi)=i\,\frac{\int d\Psi'\,C(\Psi,\Psi')\,e^{ig(\Psi')-iz\Psi'^\dagger\Psi'}}{\int d\Psi'\,e^{ig(\Psi')-iz\Psi'^\dagger\Psi'}}.
\eeq
We remark that there are visible similarities between this equation and the Bray-Rodgers equation \cite{BR, kuehn, suscaetal} derived in the context of replica method analysis of sparse random matrices. In a sense, this parallel makes (\ref{sddlpsi}) look like an equation corresponding to two bosonic and two fermionic replicas, though we do not know how to make this analogy precise.

In cases of interest, the quadratic form $\int d\Psi\,d\Psi'\, g(\Psi) C^{-1}(\Psi,\Psi')g(\Psi')$ and hence also the saddle point equation (\ref{sddlpsi}) respect a symmetry in the form of (super)rotations of $\Psi$ that preserve the inner product $\Psi_1^\dagger \Psi_2$. Under such circumstances, it is natural to look for solutions given by $g(\Psi)$ that are themselves invariant under such superrotations (an alternative would be continuous families of saddle points connected by superrotations). Such solutions must be of the form
\beq\label{gsymm}
g(\Psi)=g_*(\Psi^\dagger\Psi), \qquad \Psi^\dagger\Psi\equiv \phi^2+\chi^2+\xi\eta.
\eeq
For such functions, the saddle point equation can be simplified. It is convenient to introduce polar coordinates in the $(\phi,\chi)$-plane:
\beq\label{polar}
\rho=\phi^2+\chi^2,\qquad \phi=\sqrt{\rho}\cos\alpha,\qquad \chi=\sqrt{\rho}\sin\alpha,\qquad d\phi\,d\chi=\frac12 d\rho\, d\alpha.
\eeq

An elementary fact is that, for any function $F(\rho)$ decreasing at infinity (some of our functions oscillate at infinity but they are understood to incorporate infinitely slow asymptotic decay), one can proceed with the following derivation: first we write
\beq\label{Fsymm}
\int d\Psi F(\Psi^\dagger\Psi)=\frac12\int d\rho\, d\alpha\, d\xi\, d\eta\,  F(\rho+\xi\eta)=
\frac12\int d\rho \,d\alpha\,d\xi \,d\eta  \left[F(\rho)+F'(\rho)\xi\eta\right].
\eeq
Here, we have simply employed the Taylor expansion for $F$, which necessarily terminates at first subleading order since $\xi^2=\eta^2=0$ for anticommuting variables. Furthermore, the first term integrates to zero, while $\int d\xi\,d\eta\,\xi\eta=-1$ by the integration rules (\ref{Berezin}). 
Hence,\footnote{Equation (\ref{rotationloc}) occupies a prominent place in supersymmetry-based approaches to random matrix theory and 
has inspired a number of generalizations \cite{sup1,sup2,sup3,sup4,sup5,sup6,sup7}.}
\beq\label{rotationloc}
\int d\Psi F(\Psi^\dagger\Psi)=-\frac12\int_0^\infty\hspace{-3mm} d\rho \int_0^{2\pi}\hspace{-3mm} d\alpha\, F'(\rho)=\pi F(0).
\eeq
With this in mind, for functions of the form (\ref{gsymm}), the saddle point equation (\ref{sddlpsi}) becomes
\beq\label{sddlsymm}
g_*(\Psi^\dagger\Psi)=\frac{ie^{-ig_*(0)}}{\pi}\int d\Psi'\, C(\Psi,\Psi')\,e^{ig_*(\Psi'^\dagger\Psi')-iz\Psi'^\dagger\Psi'}.
\eeq
This equation should be further processed for each specific model depending on the form of $C$.
In all concrete cases we shall consider, (\ref{sddlsymm}) will imply\footnote{It has been kindly pointed out to us by a journal referee that $g_*(0)\!\!=\!\!0$ holds in general. Indeed, one may assume without loss of generality that $C(0,0)=0$, since an additive constant contribution in $C$ would have only affected the irrelevant overall scale in (\ref{Mav}). Furthermore, $C$ is supersymmetric and hence $C(\Psi,\Psi')=C(\Psi^\dagger\Psi,\Psi^\dagger\Psi',\Psi'^\dagger\Psi')$. Therefore, $C(0,\Psi')$ is a function of $\Psi'^\dagger\Psi'$ alone. But then, (\ref{rotationloc}) and (\ref{sddlsymm}) imply that
$g_*(0)\sim C(0,0)=0$.} 
\beq\label{gzero}
g_*(0)=0,
\eeq
which then leaves
\beq
g_*(\Psi^\dagger\Psi)=\frac{i}{\pi}\int d\Psi'\, C(\Psi,\Psi')\,e^{ig_*(\Psi'^\dagger\Psi')-iz\Psi'^\dagger\Psi'}.
\eeq
After integrating over $\xi'$ and $\eta'$, this will become an integral equation for a complex-valued function of one real variable $g_*(\rho)$.

The leading saddle point estimate for (\ref{afterMF}) is then extracted as
\beq\label{sddlapprox}
\langle(\mathbf{M}-z\mathbf{I})^{-1}_{11}\rangle\propto \frac {\int d\Psi\,(i\phi^2)\,e^{ig_*(\Psi^\dagger\Psi)-iz\Psi^\dagger\Psi}}{\int d\Psi\,e^{ig_*(\Psi^\dagger\Psi)-iz\Psi^\dagger\Psi}}e^{-N\hspace{0.1mm}S[g_*]}\,.
\eeq
The denominator is again governed by (\ref{rotationloc}) and can be evaluated as $\pi e^{ig_*(0)}=\pi$.
Furthermore, using the saddle point equation (\ref{sddlsymm}),
\beq
S[g_*]=\frac{i}{2\pi}\int d\Psi\, g_*(\Psi^\dagger\Psi)\,e^{ig_*(\Psi^\dagger\Psi)-iz\Psi^\dagger\Psi}-\ln\left(\int d\Psi\,e^{ig_*(\Psi^\dagger\Psi)-iz\Psi^\dagger\Psi}\right).
\eeq
Once again, by (\ref{rotationloc}) and (\ref{gzero}), this expression is an irrelevant $z$-independent constant. Importantly, under the assumption
that the powers of $N$ in (\ref{Mav}) are as given (and so they will be in the practical applications we shall treat below), this structure automatically explains why the distribution has a well-defined large $N$ limit as a consequence of the dominant saddle point being supersymmetric, as per (\ref{gsymm}). If, by contrast, $S[g_*]$ were $z$-dependent, no large $N$ limit would exist.

A clarification is in order here. In general, besides the factors included in (\ref{sddlapprox}), one would expect a functional determinant arising from the quadratic dependence of $S[g]$ on fluctuations of $g$ around $g_*$. In our situation, however, this determinant cannot depend on $z$, even if $g_*$ does depend on $z$, and can be safely ignored. The reason is that the saddle point $g_*$ and the functional determinant in no way depend on the presence of the insertion involving $i\phi_1^2$ and would be exactly the same without this insertion. However, without the insertion of $i\phi_1^2$, our original expression (\ref{GaussSUSY}) is exactly $z$-independent for all $N$, and hence so must be its leading saddle point estimate at large $N$, and therefore the same applies to the functional determinant in this estimate (while $S[g_*]=\mathrm{const}$ as already explained). This is consistent with the picture in \cite{MF,FM,Fyodorov}. To prove the $z$-independence of the functional determinant more directly, one would have to apply supersymmetry arguments to the integral over fluctuations, though we shall not pursue it here. Independence of supersymmetric integrals on parameters is systematically discussed in the context of `supersymmetric localization' \cite{localization,localization2}.

Putting everything together, the leading saddle point estimate for the resolvent becomes
\beq\label{resolvintPsi}
\langle(\mathbf{M}-z\mathbf{I})^{-1}_{11}\rangle\propto \int d\Psi\,(i\phi^2)\,e^{ig_*(\Psi^\dagger\Psi)-iz\Psi^\dagger\Psi}.
\eeq
Evaluating the Grassmannian integrals in a manner parallel to (\ref{rotationloc}), we get
\beq
\langle(\mathbf{M}-z\mathbf{I})^{-1}_{11}\rangle\propto i\int  d\rho\, d\alpha\, d\xi\, d\eta\, \rho\cos^2\alpha\,e^{ig_*(\rho+\xi\eta)-iz(\rho+\xi\eta)}\propto i \int_0^\infty\hspace{-3mm} d\rho\,e^{ig_*(\rho)-iz\rho} .
\eeq
The corresponding estimate for the eigenvalue distribution is expressed by (\ref{resolv}) through the imaginary part of this formula, or equivalently as
\beq\label{resolvsddl}
p(\la)\propto \Re\left[\int_0^\infty\hspace{-3mm} d\rho\,e^{ig_*(\rho)-iz\rho}\right]_{z=\la-i0}.
\eeq
In concrete cases, it will be possible to further simplify this formula using the saddle point equation (\ref{sddlsymm}). For the rest of the paper, we shall be concerned with implementing the program described above in a few concrete examples (and a slight variation of this program in the last section on normalized graph Laplacians). This amounts in practice to finding an effective way to handle the saddle point equation (\ref{sddlsymm}), which yields the eigenvalue distribution estimate (\ref{resolvsddl}).

%%%%%%%%%%%%

\subsection{An unconventional derivation of the Wigner semicircle law}\label{Wig}

As a starter, we shall re-derive the Wigner semicircle law using the formalism described above. Of course, it is something of an `overkill' to evoke the Fyodorov-Mirlin method for the Wigner semicircle, which can be alternatively derived by more elementary means. An advantage, however, is that, with all the preliminary ingredients in place, the saddle point equation (\ref{sddlsymm}) turns into an ordinary algebraic quadratic equation for one variable, whose solution was known already to the Babylonians \cite{babel}, and the semicircle pops out of this age-old solution. The steps of the Wigner semicircle derivation given here will be furthermore useful in our subsequent analysis of less obvious cases.

We shall consider $N\times N$ real symmetric matrices $\mathbf{M}$ whose diagonal entries are zero and the offdiagonal entries are i.i.d.\ random variables of variance $1/N$ and zero mean. This differs from the usual GOE ensemble in that the diagonal entries are set to zero, but this would not matter at large $N$ since there are much fewer diagonal entries than the offdiagonal ones and they do not affect the eigenvalue distribution at all, provided that the diagonal entries themselves are not too large. (In fact, it would not be difficult to incorporate nonzero diagonal entries in our derivation.) The probability distribution for
$\mathbf{M}$ is then explicitly
\beq\label{WigP}
\begin{split}
&P(\mathbf{M})=\left(\prod_i \de(M_{ii})\right)\prod_{i<j} \left[\sqrt{N}h(\sqrt{N}M_{ij})\,\de(M_{ji}-M_{ij})\right],\\
&\int dx h(x)=1,\qquad  \int dx \,x h(x)=0,\qquad  \int\, dx\, x^2 h(x)=1.
\end{split}
\eeq
We then have to average the resolvent formula (\ref{Gausssuperv}) over this ensemble.
This yields
\beq\label{GaussWig}
\langle(\mathbf{M}-z\mathbf{I})^{-1}_{11}\rangle\propto \int d\pmb{\Psi}\,\phi_1^2\, e^{\sum_{k<l}\ln[\tilde h(2 \Psi^\dagger_k \Psi_l/\sqrt{N})]-iz\sum_k \Psi^\dagger_k\Psi_k},
\eeq
where $\tilde h(k)\equiv\int dx\,e^{ikx}h(x)$ is the Fourier transform of $h$. We can expand $\tilde h$ in a Taylor series, with the first few terms fixed by (\ref{WigP}) as $\tilde h(k)=1-k^2/2+O(k^3)$. The higher terms are suppressed in (\ref{GaussWig}) by higher powers of $N$ and drop out in the large $N$ limit as a reflection of the Wigner universality \cite{MF}. Then, expanding the logarithm as well, we end up with\footnote{\label{diagnegl}The summation is, strictly speaking, over $k\ne l$. We have extended it to all $k$ and $l$, which in principle requires appending a compensating factor of $\exp[\sum_k (\Psi^\dagger_k\Psi_k)^2/N]$. This expression, however, first, contains $1/N$ suppression and, second, factorizes over the components of $\pmb{\Psi}$. Hence, after the Fyodorov-Mirlin transform (\ref{MFtransfrm}) has been applied and the entire integrand of $d\pmb{\Psi}$ factorizes over the components of $\pmb{\Psi}$, this extra factor that we have omitted would only give tiny contributions, suppressed by $1/N$, to each integral over a component of $\pmb{\Psi}$, and can thus be ignored relative to the contributions that we are keeping explicitly.}
\beq\label{GaussWigC}
\langle(\mathbf{M}-z\mathbf{I})^{-1}_{11}\rangle\propto \int d\pmb{\Psi}\,\phi_1^2\, \exp\left[-\frac1{N}\sum_{kl}(\Psi^\dagger_k \Psi_l)^2-iz\sum_k \Psi^\dagger_k\Psi_k\right],
\eeq
This is manifestly of the form (\ref{Mav}), as promised, with $C(\Psi,\Psi')\equiv 2(\Psi^\dagger\Psi')^2$. We can then fastforward through the application of the Fyodorov-Mirlin method until we reach the saddle point equation in the form (\ref{sddlsymm}) written as
\beq\label{sddlWig}
g_*(\Psi^\dagger\Psi)=\frac{2ie^{-ig_*(0)}}{\pi}\int d\Psi'\,(\Psi^\dagger\Psi')^2\,e^{ig_*(\Psi'^\dagger\Psi')-iz\Psi'^\dagger\Psi'}
\eeq
First, substituting $\Psi=0$ immediately leads to the conclusion that $g_*(0)=0$ as in (\ref{gzero}). To simplify the integral over $\Psi'$, we introduce polar coordinates as in (\ref{polar}) for both $\Psi$ and $\Psi'$, so that $\alpha'$ is the angle between the 2-vectors $(\phi,\chi)$ and $(\phi',\chi')$. One then has
$$
\int d\Psi'\,(\Psi^\dagger\Psi')^2\,e^{ig_*-iz\Psi'^\dagger\Psi'}
=\frac12 \int d\rho'd\alpha'd\xi'd\eta'\left[\sqrt{\rho\rho'}\cos\alpha' +\frac{\xi\eta'+\eta\xi'}2\right]^2\! e^{ig_*(\rho'+\xi'\eta')-iz(\rho'+\xi'\eta')}.
$$
We can focus on the part of this expression that does not depend on $\xi$ and $\eta$ since $g_*$ only depends on $\rho+\xi\eta$, and supersymmetry guarantees  that, if (\ref{sddlWig}) is satisfied among the purely commutative terms, it is satisfied exactly. (The Grassmannian components of the saddle point equation for a related sparse random matrix problem can be seen explicitly in \cite{FM}.) Extracting these $(\xi,\eta)$-independent terms and treating the integration over $\xi'$, $\eta'$ and $\alpha'$ as in  (\ref{Fsymm}-\ref{rotationloc}), we obtain
$$
\int d\Psi'\,(\Psi^\dagger\Psi')^2\,e^{ig_*-iz\Psi'^\dagger\Psi'}
\to -\frac{\rho}2 \int d\rho'd\alpha'\, \rho'\cos^2\alpha' \frac{\del}{\del\rho'}e^{ig_*(\rho')-iz\rho'}
=\frac{\pi\rho}2\int d\rho'\,e^{ig_*(\rho')-iz\rho'}.
$$
Substituting all of this back into (\ref{sddlWig}), we get
\beq\label{gstWig}
g_*(\rho)={i\rho}\int_0^\infty\hspace{-3mm} d\rho'\,e^{ig_*(\rho')-iz\rho'}.
\eeq
Thus, $g_*$ is simply proportional to $\rho$ and can hence be written as $g_*(\rho,z)=-\rho W(z)$. Taking the integral, we get
\beq\label{W}
-W=\frac{1}{W+z},\qquad W^2+zW+{1}=0,\qquad W=\frac{-z-i\sqrt{4-z^2}}{2}.
\eeq
Finally, from (\ref{resolvsddl}) and (\ref{gstWig}),
\beq\label{semicircle}
p_W(\la)\propto \Re\left[\int_0^\infty\hspace{-3mm} d\rho\,e^{ig_*(\rho)-iz\rho}\right]_{z=\la}=\Re\left[\frac{2g_*(\rho,\la)}{i\rho}\right]\propto \Im[W(\la)]\propto \sqrt{4-\la^2},
\eeq
for $|\la|<2$, and 0 otherwise.
This is precisely in the form of the Wigner semicircle (and the width agrees with GOE \cite{randmat} scaled so that the variance of the offdiagonal entries is $1/N$).

%%%%%%%%%%%%%%%%%%%%%%%%%%%%%%%%%%%%%%%

\section{Dense zero-row-sum matrices}

\subsection{The Fyodorov distribution}\label{Fyo}

We now turn to the much-less-trivial and apparently less-widely-appreciated case of zero-row-sum matrices.
Our review will generally follow the original derivations of \cite{Fyodorov}, though using more elementary and hopefully more accessible notation. This case is not straightforwardly tractable with more standard random matrix methods. Following the original treatment in \cite{Fyodorov}, the distribution was re-derived in \cite{SMFL} using the method of moments, though it requires non-obvious resummations. (See \cite{moments} for a more recent application of the method of moments to sparse random matrix problems.) The Fyodorov-Mirlin method, as originally employed for this problem in \cite{Fyodorov}, on the other hand, develops an analytic theory of this eigenvalue distribution systematically. 

The ensemble is very similar to section~\ref{Wig}, except that now the diagonal entries are nonzero and filled in precisely in such a way as to make the row sums vanish:
\beq\label{zsum}
\forall i:\quad M_{ii}=-\sum_{j\ne i} M_{ij}.
\eeq
The joint probability distribution of the entries of the real symmetric matrix $\mathbf{M}$ is
\beq\label{FyoP}
\begin{split}
&P(\mathbf{M})=\left(\prod_i \de({\textstyle \sum_j} M_{ij})\right)\prod_{i<j} \left[\sqrt{N}h(\sqrt{N}M_{ij})\,\de(M_{ji}-M_{ij})\right],\\
&\int dx h(x)=1,\qquad  \int dx \,x h(x)=0,\qquad  \int\, dx\, x^2 h(x)=1.
\end{split}
\eeq
This apparently minor change away from (\ref{WigP}) has a significant effect on the large $N$ eigenvalue distribution.

In view of (\ref{zsum}) and the symmetry of $\mathbf{M}$, (\ref{Gausssuperv}) can be written as
\beq\label{Fyores}
(\mathbf{M}-z\mathbf{I})^{-1}_{11}\propto \int d\pmb{\Psi}\,\phi_1^2\, e^{-i\sum_{k<l}M_{kl} (\Psi_k- \Psi_l)^\dagger (\Psi_k- \Psi_l)-iz\sum_k \Psi^\dagger_k\Psi_k}.
\eeq
The components $M_{kl}$ with $k<l$ are all independent, while the integrand has been factorized in terms of these components. 
The averaging over $\mathbf{M}$ then amounts to evaluating, for each given $k<l$,
$$
\int dM_{kl}\,\sqrt{N}\,h(\sqrt{N}M_{kl})\,  e^{-i M_{kl} (\Psi_k- \Psi_l)^\dagger (\Psi_k- \Psi_l)}=\tilde h[ -(\Psi_k- \Psi_l)^\dagger (\Psi_k- \Psi_l)/\sqrt{N}],
$$
with $\tilde h(k)\equiv\int dx\, e^{ikx}h(x)$. Putting the contributions from all $k<l$ together yields
\beq\label{Fyoresav}
\langle(\mathbf{M}-z\mathbf{I})^{-1}_{11}\rangle\propto \int d\pmb{\Psi}\,\phi_1^2\, e^{\sum_{k<l}\ln\{\tilde h[ -(\Psi_k- \Psi_l)^\dagger (\Psi_k- \Psi_l)/\sqrt{N}]\}-iz\sum_k \Psi^\dagger_k\Psi_k}.
\eeq
Expanding the Fourier transform $\tilde h$ through the moments of $h$, as done under (\ref{GaussWig}), and dropping the higher terms suppressed by higher powers of $1/N$, we end up with
\beq\label{FyoC}
\langle(\mathbf{M}-z\mathbf{I})^{-1}_{11}\rangle\propto \int d\pmb{\Psi}\,\phi_1^2\, \exp\left[-\frac1{4N}\sum_{kl}[(\Psi_k- \Psi_l)^\dagger (\Psi_k- \Psi_l)]^2-iz\sum_k \Psi^\dagger_k\Psi_k\right],
\eeq
which is again of the form (\ref{Mav}) with $C(\Psi,\Psi')\equiv [(\Psi- \Psi')^\dagger (\Psi- \Psi')]^2/2$. As in section~\ref{Wig}, we fastforward with an application of the Fyodorov-Mirlin method till we reach the saddle point equation  (\ref{sddlsymm}), which takes the form \cite{Fyodorov}
\beq\label{sddlFyo}
g_*(\Psi^\dagger\Psi)=\frac{ie^{-ig_*(0)}}{2\pi}\int d\Psi'\, [(\Psi- \Psi')^\dagger (\Psi- \Psi')]^2\,e^{ig_*(\Psi'^\dagger\Psi')-iz\Psi'^\dagger\Psi'}.
\eeq
Note that, when $\Psi=0$, the right-hand-side vanishes by (\ref{rotationloc}), and hence $g_*(0)=0$ as in (\ref{gzero}). It remains to introduce polar coordinates (\ref{polar}) for $\Psi$ and $\Psi'$ and evaluate the integrals over $\xi'$, $\eta'$ and $\alpha'$ in a manner analogous to the derivations under (\ref{sddlWig}).
As before, $g_*$ is completely fixed by looking at the terms in (\ref{sddlFyo}) independent of $\xi$ and $\eta$ (and the remaining terms must match on the two sides automatically due to the symmetry with respect to superrotations). We then have
\begin{align*}
&\int d\Psi'\, [(\Psi- \Psi')^\dagger (\Psi- \Psi')]^2\,e^{ig_*(\Psi'^\dagger\Psi')-iz\Psi'^\dagger\Psi'}\\
&=\frac12\int d\rho'\,d\alpha'\,d\xi'\,d\eta'\,[\rho+\rho'-2\sqrt{\rho\rho'}\cos\alpha'+(\xi-\xi')(\eta-\eta')]^2\left(1+\xi'\eta'\del_{\rho'}\right)e^{ig_*(\rho')-iz\rho'}.
\end{align*}
Then, with $\int d\xi' d\eta' \xi'\eta'=-1$ and some integration by parts including the boundary terms at $\rho'=0$, retaining only the terms independent of $\xi$ and $\eta$, we get
\beq
\int d\Psi'\, [(\Psi- \Psi')^\dagger (\Psi- \Psi')]^2\,e^{ig_*(\Psi'^\dagger\Psi')-iz\Psi'^\dagger\Psi'}\to \pi\rho^2+2\pi\rho\int_0^\infty\hspace{-3mm} d\rho'e^{ig_*(\rho')-iz\rho'}.
\eeq
Substituting this back into (\ref{sddlFyo}), we get
\beq\label{Fyog}
g_*(\rho)=\frac{i\rho^2}2+i\rho\int_0^\infty\hspace{-3mm}  d\rho'e^{ig_*(\rho')-iz\rho'}.
\eeq
Then $g_*$ is manifestly a quadratic polynomial of $\rho$ conveniently parametrized as
\beq\label{Fyogpoly}
g_*(\rho,z)=\frac{i\rho^2}2+(iF(z)+z)\rho.
\eeq 
Substituting this expression into (\ref{Fyog}) yields an integral equation for $F$ originally derived in an equivalent form in \cite{Fyodorov}:
\beq\label{Feq}
F(z)-iz=\int_0^\infty\hspace{-3mm}  d\rho\, e^{-F(z)\rho-\rho^2/2}.
\eeq
This integral equation is not too pleasant to work with in practice, but it is easily converted into a differential  equation by differentiating with respect to $z$ and then writing $\rho e^{-F(z)\rho-\rho^2/2}=-(F+\del_\rho) e^{-F(z)\rho-\rho^2/2}$, which yields
\beq\label{FyoODE}
\frac{dF}{dz}=\frac{i}{2+izF-F^2}.
\eeq
This equation has to be solved\footnote{Note that (\ref{FyoODE}) can be integrated explicitly to yield 
$$iz(F)=F-F(0)\,e^{[F^2-F^2(0)]/2}+e^{F^2/2}\int_{F(0)}^F e^{-x^2/2}dx,$$ 
which defines $F(z)$ implicitly. In practice, this representation is not of much use however, and the most straightforward way to recover the curve $F(z)$ is by solving (\ref{FyoODE}).} with the initial condition
\beq\label{Fini}
F(0)=f_0,\qquad f_0=\int_0^\infty\hspace{-3mm}  d\rho\, e^{-f_0\rho-\rho^2/2}\approx 0.751791.
\eeq
Finally, from (\ref{Fyog}) and (\ref{Fyogpoly}), $\int d\rho'e^{ig_*(\rho')-iz\rho'}=F(z)-iz$, and then, from (\ref{resolvsddl}) we obtain a formula for the Fyodorov distribution that governs the eigenvalues of symmetric zero-row-sum random matrices:
\beq\label{pzrs}
p_{\mathrm{zrs}}(\la)\propto \Re[F(\la)].
\eeq
\begin{figure}[t]
\centering
\begin{subfigure}{0.45\textwidth}
\hspace{-1cm}\includegraphics[width = 1.2\linewidth]{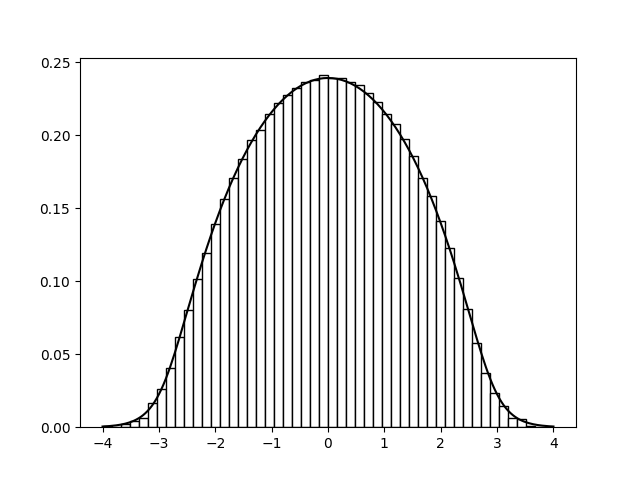}\vspace{-8mm}
\begin{picture}(0,0)
\put(8,147){$p_{\mathrm{zrs}}$}
\put(190,18){$\la$}
\end{picture}
\caption{\rule{0.1cm}{0cm}}
\end{subfigure}
\begin{subfigure}{0.45\textwidth}
\hspace{-3mm}\includegraphics[width = 1.2\linewidth]{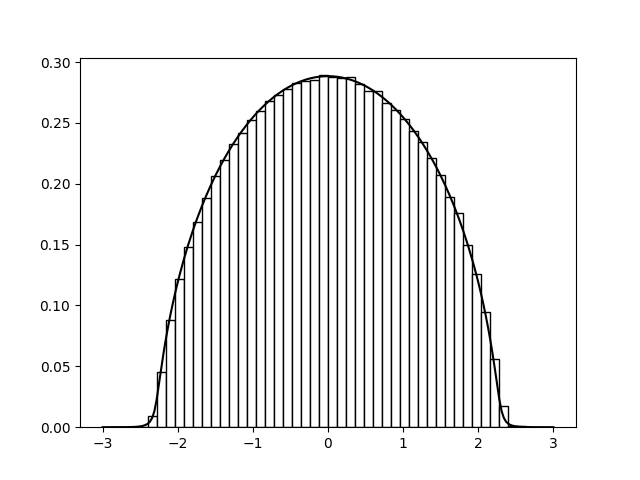}\vspace{-8mm}
\begin{picture}(0,0)
\put(28,147){$p_{\mathrm{\mu=1/2}}$}
\put(208,18){$\la$}
\end{picture}
\caption{\rule{-1.3cm}{0cm}}
\end{subfigure}
\caption{Eigenvalue distributions of individual instances of $10000\times 10000$ random matrices with normally distributed independent entries: (a) the matrix is drawn from the ensemble (\ref{FyoP}), which is the same as (\ref{muP}) at $\mu=1$, and matched against the properly normalized Fyodorov distribution (\ref{pzrs}); (b) the matrix is drawn from the ensemble (\ref{muP}) with $\mu=1/2$ and matched against the properly normalized analytic curve (\ref{pmu}). Evidently, the edges of the distribution get sharper and the top flatter as one decreases $\mu$ and approaches the Wigner semicircle at $\mu=0$.}
\label{fig:dense}
\end{figure}
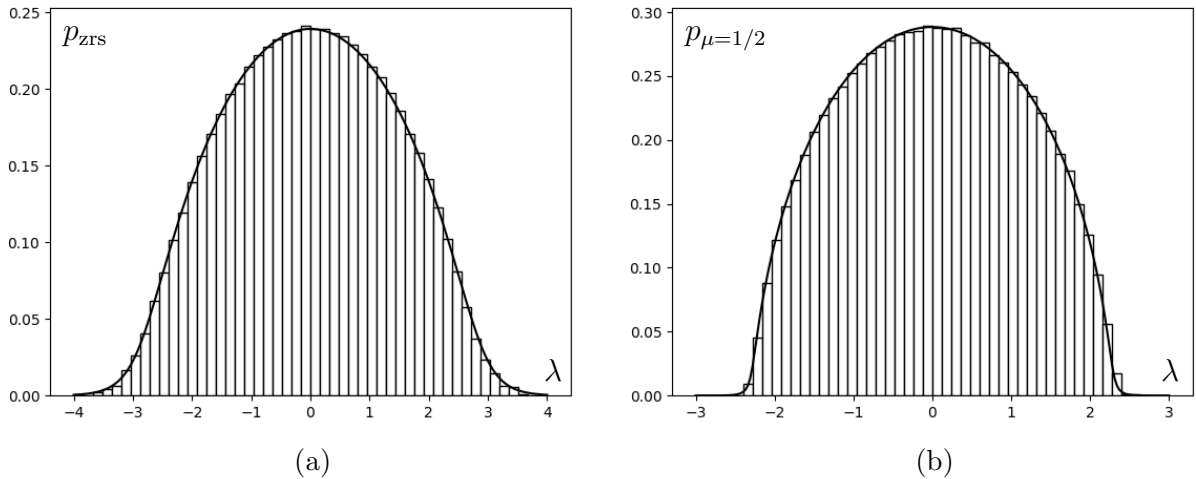
\!\!\!\! In Fig.~\ref{fig:dense}a, we display the Fyodorov distribution $p_{\mathrm{zrs}}$ computed from (\ref{FyoODE}-\ref{pzrs}) and normalized to one and the corresponding numerical simulation of the ensemble 
(\ref{FyoP}), showing perfect agreement.

%%%%%%%%%%%%

\subsection{An interpolating ensemble}

It is a natural thought that the curve in Fig.~\ref{fig:dense}a looks like a cross between the Wigner semicircle and a Gaussian. This idea can be made more precise by constructing an explicit ensemble whose eigenvalue density interpolates between the semicircle and a Gaussian, with the zero-row-sum ensemble (\ref{FyoP}) appearing as an intermediate point.

An intuitive picture why the zero-row-sum ensemble deviates from the Wigner universality class is that the diagonal entries are much larger than the offdiagonal ones. Then, the presence of these large diagonal entries strongly affects the diagonalization process. In \cite{zrs}, the situation was compared to the Rosenzweig-Porter (or `Brownian') ensemble, where the diagonal and offdiagonal entries are independent, but their variance is different. This analogy seems to suggest the correct picture, but it would take extra effort to make the connection precise, which has in fact been accomplished in the mathematical literature on free probability \cite{free}.

To elucidate the situation directly, we propose instead to consider a generalization of the condition (\ref{zsum}) given by
\beq\label{musum}
\forall i:\quad M_{ii}=-\mu \sum_{j\ne i} M_{ij},
\eeq
where $\mu$ is a parameter. When $\mu=1$, we are back to (\ref{FyoP}). When $\mu=0$, it is exactly the ensemble treated in section~\ref{Wig}, whose eigenvalue distribution is given by the Wigner semicircle. Finally, when $\mu\to\infty$, the offdiagonal entries become negligible, and the eigenvalue distribution is simply the distribution of the diagonal entries, which is Gaussian by the central limit theorem.

For the matrix entry distribution
\beq\label{muP}
\begin{split}
&P(\mathbf{M})=\left(\prod_i \de(M_{ii}+\mu{\textstyle \sum_{j\ne i}} M_{ij})\right)\prod_{i<j} \left[\sqrt{N}h(\sqrt{N}M_{ij})\,\de(M_{ji}-M_{ij})\right],\\
&\int dx h(x)=1,\qquad  \int dx \,x h(x)=0,\qquad  \int\, dx\, x^2 h(x)=1,
\end{split}
\eeq
the derivations given above can be repeated with minimal modifications, yielding an analytic theory for the eigenvalue distributions of these $\mu$-ensembles. We summarize below a few crucial intermediate steps and the final result.

For the $\mu$-ensemble, (\ref{Gausssuperv}) becomes
\beq\label{mures}
(\mathbf{M}-z\mathbf{I})^{-1}_{11}\propto \int d\pmb{\Psi}\,\phi_1^2\, e^{i\sum_{k<l}M_{kl} [2\Psi_k^\dagger \Psi_l-\mu(\Psi^\dagger_k\Psi_k+\Psi^\dagger_l\Psi_l)]-iz\sum_k \Psi^\dagger_k\Psi_k}.
\eeq
Following through with the usual steps after the $\mathbf{M}$-averaging, we end up with a representation of the form (\ref{Mav}) with $C(\Psi,\Psi')=[2\Psi^\dagger \Psi'-\mu(\Psi^\dagger\Psi+\Psi'^\dagger\Psi')]^2/2$.
This yields a saddle point equation (\ref{sddlsymm}) in the form
\beq\label{sddlmu}
g_*(\Psi^\dagger\Psi)=\frac{ie^{-ig_*(0)}}{2\pi}\int d\Psi'\,[\mu(\Psi^\dagger\Psi+\Psi'^\dagger\Psi')-2\Psi^\dagger \Psi']^2\,e^{ig_*(\Psi'^\dagger\Psi')-iz\Psi'^\dagger\Psi'}.
\eeq
Once again, $g_*(0)$ manifestly vanishes by (\ref{rotationloc}), and the Grassmannian integrals can be performed to yield an integral equation for an ordinary function:
\beq\label{mug}
g_*(\rho)=\frac{i\mu^2\rho^2}2+i\rho\int_0^\infty\hspace{-3mm}  d\rho'e^{ig_*(\rho')-iz\rho'},
\eeq
cf.\ (\ref{gstWig}) and (\ref{Fyog}). We can look for a solution in the form
\beq\label{mupoly}
g_*(\rho,z)=\frac{i\mu^2\rho^2}2+(iF_\mu(z)+z)\rho, 
\eeq 
so that
\beq
F_\mu(z)-iz=\int_0^\infty\hspace{-3mm}  d\rho\, e^{-F_\mu(z)\rho-\mu^2\rho^2/2}.
\eeq
This is converted to an ODE as
\beq\label{muODE}
\frac{dF_\mu}{dz}=\frac{i\mu^2}{1+\mu^2+izF_\mu-F_\mu^2},\qquad F_\mu(0)=\int_0^\infty\hspace{-3mm}  d\rho\, e^{-F_\mu(0)\rho-\mu^2\rho^2/2}.
\eeq
The eigenvalue density (\ref{resolvsddl}) is then
\beq\label{pmu}
p_\mu (\la)\propto \Re[F_\mu(\la)].
\eeq
In Fig.~\ref{fig:dense} we give a properly normalized plot of this curve matched against the numerics for the case $\mu=1/2$ that lies half-way between the ensembles of sections~\ref{Wig} and \ref{Fyo}, showing perfect agreement.

%%%%%%%%%%%%%%%%%%%%%%%%%%%%%%%%%%%%%%%

\section{Eigenvalues of graph Laplacians}

Having reviewed the Fyodorov-Mirlin method \cite{FM,MF} and Fyodorov's solution for the eigenvalue distribution of zero-row-sum matrices \cite{Fyodorov}, and additionally proposed an explicit family of random matrix ensembles with linear row constraints whose eigenvalue density interpolates between the Wigner semicircle and a Gaussian, we now turn to our main topic: sparse matrices and graph Laplacians.
Instead of (\ref{FyoP}), we shall consider symmetric $N\times N$ zero-row-sum matrices with each row having on average $c$ nonzero offdiagonal entries, with these entries having an arbitrary prescribed probability distribution $h$. (Technically, we will assume that the Fourier transform of $h$ is sufficiently differentiable at the origin, which requires the existence of its moments.) The joint probability of the matrix entries is written as
\beq\label{zrssparse}
P(\mathbf{M})=\left(\prod_i \de({\textstyle \sum_j} M_{ij})\right)\prod_{i<j} \left\{\left[\left(1-\frac{c}{N}\right)\de(M_{ij})+\frac{c}{N}h(M_{ij})\right]\de(M_{ji}-M_{ij})\right\},
\eeq
with $\int dx\,h(x)=1$. We have to repeat the considerations of section~\ref{Fyo} for this ensemble and analyze the emerging saddle point equation. We shall mostly focus on the regime $1\ll c\ll N$, meaning that $N$ is effectively sent to $\infty$ right away with all $1/N$ corrections discarded. On the other hand, $c$ will be treated as a tunable large parameter (in practice, of order 10 or 100) and we will develop results valid asymptotically at large $c$ and track down the quality of approximation at smaller values of $c$.

A special choice that is of central importance for us is 
\beq\label{hER}
h(x)=\de(x+1), 
\eeq
in which case $\mathbf{M}$ is literally the ordinary Laplacian (\ref{ordLa}) of an \ER random graph with $N$ vertices and the edges between any pair of vertices filled randomly and independently with probability $c/N$ so that the mean
vertex degree at large $N$ is $c$. In that case, the offdiagonal elements of $\mathbf{L}$ given by (\ref{ordLa}) are -1 with probability $c/N$ and 0 with probability $1-c/N$, while the diagonal entries are filled in to ensure that all rows sum to zero. This precisely corresponds to (\ref{zrssparse}) with $h$ given by (\ref{hER}).

\subsection{The saddle point equation}\label{sddlLap}

Averaging the resolvent representation (\ref{Gausssuperv}) over the ensemble (\ref{zrssparse}) yields
\beq\label{sparseresav}
\langle(\mathbf{M}-z\mathbf{I})^{-1}_{11}\rangle\propto \int d\pmb{\Psi}\,\phi_1^2\, e^{\sum_{k<l}\ln\left\{1+\frac{c}N\left(\tilde h[ -(\Psi_k- \Psi_l)^\dagger (\Psi_k- \Psi_l)]-1\right)\right\}-iz\sum_k \Psi^\dagger_k\Psi_k},
\eeq
with $\tilde h(k)\equiv \int dx\, e^{ikx} \,h(x)$. Assuming $c\ll N$, we can approximate the logarithm as
\beq\label{sparseexpnd}
\langle(\mathbf{M}-z\mathbf{I})^{-1}_{11}\rangle\propto \int d\pmb{\Psi}\,\phi_1^2\, e^{\frac{c}{2N}\sum_{kl}\left(\tilde h[ -(\Psi_k- \Psi_l)^\dagger (\Psi_k- \Psi_l)]-1\right)-iz\sum_k \Psi^\dagger_k\Psi_k}.
\eeq
This is manifestly of the form (\ref{Mav}) with $C(\Psi,\Psi')\equiv c\big(1-\tilde h[ -(\Psi- \Psi')^\dagger (\Psi- \Psi')]\big)$ so that the Fyodorov-Mirlin transformation can be applied
yielding the saddle point equation (\ref{sddlsymm}) in the form
\beq\label{sparsesddl}
g_*(\Psi^\dagger\Psi)=-\frac{ic\,e^{-ig_*(0)}}{\pi}\int d\Psi'\, \big(\tilde h[ -(\Psi- \Psi')^\dagger (\Psi- \Psi')]-1\big)\,e^{ig_*(\Psi'^\dagger\Psi')-iz\Psi'^\dagger\Psi'}.
\eeq
Once again, $g_*(0)=0$ by virtue of (\ref{rotationloc}) and $\tilde h(0)=1$.

As in the previous examples, to process the saddle point equation further, we write out the supervectors $\Psi$ and $\Psi'$ through their components (\ref{Psicomp}) and introduce polar coordinates (\ref{polar}) for the commuting components to obtain
\begin{align*}
&\int d\Psi'\, \big(\tilde h[ -(\Psi- \Psi')^\dagger (\Psi- \Psi')]-1\big)\,e^{ig_*(\Psi'^\dagger\Psi')-iz\Psi'^\dagger\Psi'}\\
&=\frac12 \int d\rho' d\alpha' d\xi' d\eta' dx\,h(x)\left(e^{-ix(\rho+\rho'-2\sqrt{\rho\rho'}\cos\alpha'+(\xi-\xi')(\eta-\eta'))}-1\right)\left(1+\xi'\eta'\del_{\rho'}\right)e^{ig_*(\rho')-iz\rho'}.
\end{align*}
Retaining only the terms independent of $\xi$ and $\eta$ and evaluating all integrals, we obtain
\beq\label{sddlLgen}
g_*(\rho)=-ic\left\{\tilde h(-\rho)-1-\int_0^\infty\hspace{-3mm}d\rho'e^{ig_*(\rho')-iz\rho'}\sqrt{\frac{\rho}{\rho'}}\int dx\, x\,h(x)\,e^{-ix(\rho+\rho')}J_1(2x\sqrt{\rho\rho'})\right\},
\eeq
with the Bessel function $J_1(y)\equiv (2\pi i)^{-1}\int_0^{2\pi}e^{iy\cos\theta}\cos\theta\,d\theta$.
This equation can be compared in its structure, first, to the analogous equation (\ref{Fyog}) for dense zero-row-sum matrices (and the relation between the two will become clearer as we proceed), and second, to the analogous equation for sparse matrices without row constraints in \cite{MF} that likewise features $J_1$.

As already emphasized, we are specifically interested in the case $h(x)=\de(x+1)$, $\tilde h(k)=e^{-ik}$ that corresponds to ordinary Laplacians of \ER random graphs. The saddle point equation is then written as
\beq\label{sddlLER}
g_*(\rho)=-ic\left\{e^{i\rho}-1-\rho\,e^{i\rho}\int_0^\infty\hspace{-3mm}d\rho'\,\frac{J_1(2\sqrt{\rho\rho'})}{\sqrt{\rho\rho'}}e^{ig_*(\rho')-i(z-1)\rho'}\right\}.
\eeq

The eigenvalue density should be computed from (\ref{resolvsddl}). For $g_*$ satisfying (\ref{sddlLER}), this expression can be simplified to
\beq\label{pL}
p_L(\lambda)\propto \Im[\del_\rho g_*(\rho,z)]\Big|_{\rho=0,\,\,z=\lambda+1}.
\eeq
Note that the shift of $\la$ by $+1$ on the right-hand side is specific to the saddle-point estimate of the eigenvalue distribution of the \ER graph Laplacian, and does not occur, for example, for the related case of sparse matrices without row constraints treated in \cite{MF}. (A recent mathematical
discussion on the convergence of the eigenvalue distributions to the deterministic limit for a broad class of Laplacian matrices, including our current cases of interest, can be found in
\cite{heavysparse}.)

It would be exciting to analyze equation (\ref{sddlLER}) in more detail, but it does not appear tractable at first sight. We can nonetheless effectively study it in the large $c$ regime, producing an analytic theory of the distribution curve (\ref{pL}).

%%%%%%%%%%%%

\subsection{The large $c$ expansion}\label{largecLap}

To elucidate the large $c$ behavior of (\ref{sddlLER}), we introduce the following redefinitions:
\beq\label{scaleL}
z=c+1+{\tilde z}{\sqrt{c}},\qquad \rho=\frac{\tilde \rho}{\sqrt{c}},\qquad g_*(\rho(\tilde\rho),z(\tilde z))=\tilde g(\tilde\rho,\tilde z)+\tilde\rho\sqrt{c}.
\eeq
We will drop all the tildes until the end of this section so as not to clutter the formulas. Keeping in mind that $J_1(2y)=y+O(y^3)$, in the new notation, (\ref{sddlLER}) becomes
\beq\label{gLexpand}
g=\frac{i\rho^2}2-\frac{\rho^3}{6\sqrt{c}}+i\rho\left(1+\frac{i\rho}{\sqrt{c}}\right)
\int_0^\infty\hspace{-3mm}d\rho'e^{ig_*(\rho')-iz\rho'}+O(1/c).
\eeq
At large $c$, this equation converges to (\ref{Fyog}), solved by (\ref{Fyogpoly}). The corrections are organized as a power series in $1/\sqrt{c}$, and we will only keep the first subleading order.
We can look for a solution to (\ref{gLexpand}) in the form
\beq\label{gLpoly}
g=\left(iF(z)+z+\frac{if(z)}{\sqrt{c}}\right)\rho+\left(\frac{i}2+\frac{v(z)}{\sqrt{c}}\right)\rho^2-\frac{\rho^3}{6\sqrt{c}},
\eeq
where $F(z)$ satisfies (\ref{Feq}-\ref{Fini}). Substituting this expression into (\ref{gLexpand}), the order $O(1)$ is satisfied by construction, while at order $O(1/\sqrt{c})$ we get
\beq
f=\int_0^\infty\hspace{-3mm}d\rho\,e^{-F\rho-\rho^2/2}\left(-f\rho+iv\rho^2-i\rho^3/6\right),\qquad 
v=-\int_0^\infty\hspace{-3mm}d\rho\,e^{-F\rho-\rho^2/2}=-(F-iz).
\eeq
Hence,
\beq\label{fratio}
f=-\frac{i}6\,\frac{6(F-iz)\int d\rho\,\rho^2\,e^{-F\rho-\rho^2/2}+\int d\rho\,\rho^3\,e^{-F\rho-\rho^2/2}}{1+\int d\rho\,\rho^2\,e^{-F\rho-\rho^2/2}}.
\eeq
Finally, using (\ref{Feq}) and (\ref{FyoODE}), we can write
\beq
\int_0^\infty\hspace{-3mm} d\rho\,\rho^n\,e^{-F\rho-\rho^2/2}=\left(-\frac{1}{\del_z F}\del_z\right)^n\int_0^\infty\hspace{-3mm} d\rho\,e^{-F\rho-\rho^2/2}=(i(2+izF-F^2)\del_z)^n\,(F-iz).
\eeq
Applying this relation and expressing all $z$ derivatives of $F$ through $F$ and $z$ using (\ref{FyoODE}),
we obtain an explicit formula for $f$:
\beq\label{fexpl}
f=-\frac{i}6\,\frac{5F^4-11izF^3-6z^2F^2-2F^2-6z^2-3izF+2}{2+izF-F^2}.
\eeq
Finally, with the rescaling (\ref{scaleL}) taken into account, we get from (\ref{pL})
\beq\label{pLsqrtc}
p_L(\la)\Big|_{\la=c+\sqrt{c}z}\propto \Re\left[F(z)+\frac{f(z)}{\sqrt{c}}\right],
\eeq
with $F$ and $f$ effectively recovered from (\ref{FyoODE}-\ref{Fini}) and (\ref{fexpl}). Evidently, at very large $c$, the second term can be ignored, and the curve tends to the Fyodorov distribution $p_{\mathrm{zrs}}$, given by (\ref{pzrs}), re-centered at $c$ and scaled by $\sqrt{c}$.

\begin{figure}[t]
\centering
\begin{subfigure}{0.45\textwidth}
\hspace{-1.3cm}\includegraphics[width = 1.2\linewidth]{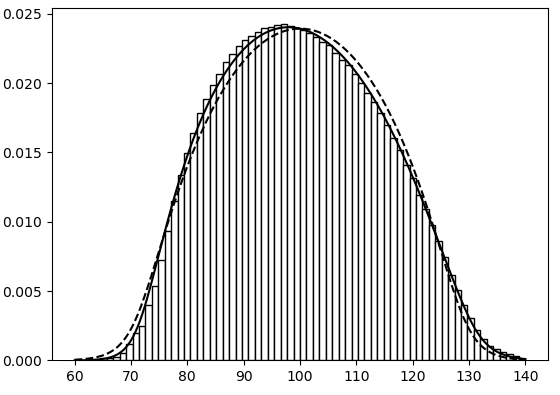}\vspace{-4mm}
\begin{picture}(0,0)
\put(-8,165){$p_L$}
\put(170,165){$c=100$}
\put(200,23){$\la$}
\end{picture}
\caption{\rule{0.3cm}{0cm}}
\end{subfigure}
\begin{subfigure}{0.45\textwidth}
\hspace{1mm}\includegraphics[width = 1.2\linewidth]{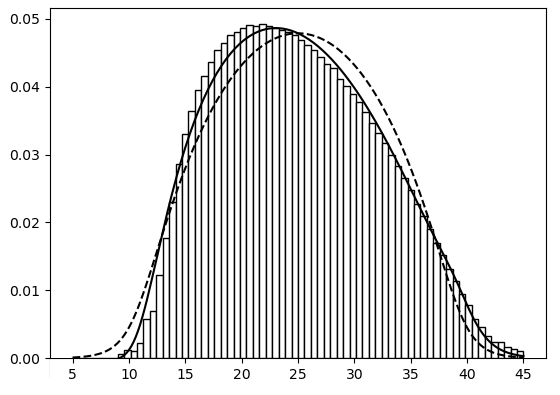}\vspace{-4mm}
\begin{picture}(0,0)
\put(30,165){$p_L$}
\put(240,25){$\la$}
\put(215,165){$c=25$}
\end{picture}
\caption{\rule{-2.5cm}{0cm}}
\end{subfigure}
\caption{Theoretical prediction (\ref{pLsqrtc}) plotted as solid lines compared with the eigenvalue density of the Laplacian
(\ref{ordLa}) of an \ER random graph with $N=25000$ vertices and mean degree (a) $c=100$ and (b) $c=25$. The dashed lines represent the shifted and scaled Fyodorov distribution (\ref{pzrs}) that corresponds to discarding the $1/\sqrt{c}$ correction in (\ref{pLsqrtc}).}
\label{figL}
\end{figure}
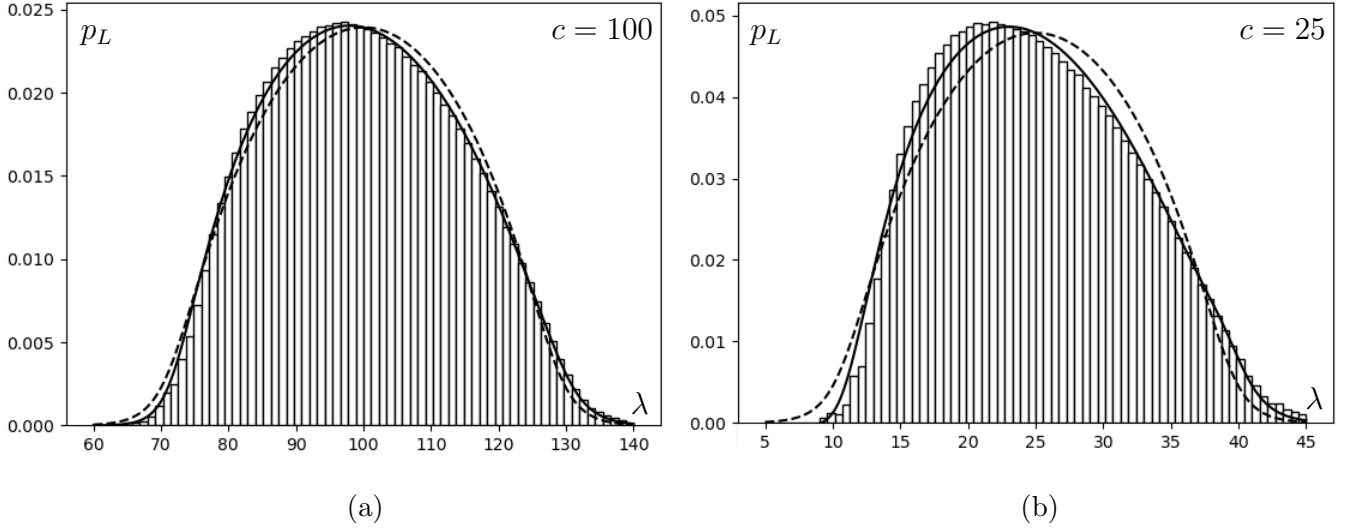
We have provided a comparison of this theoretical prediction with numerical sampling in Fig~\ref{figL}. (The elementary Python code we have used is quoted in the Appendix.) At $c=100$, the agreement is essentially perfect. At $c=25$ the shape of the curve remains correct, though there are visible deviations at the level of a few percent, as one would naturally expect from the structure of the $1/\sqrt{c}$ expansion. 

%%%%%%%%%%%%%%%%%%%%%%%%%%%%%%%%%%%%%%%

\section{Normalized graph Laplacians}\label{secnorm}

For the normalized graph Laplacian (\ref{normLa}), the computational routine reviewed and developed above has to be somewhat modified, also taking us beyond the established settings of \cite{MF,FM,Fyodorov}.
These modifications, however, do not affect the general principles that lead to the saddle point estimates in the previous sections. We now proceed with an explicit analysis of this case.

\subsection{Auxiliary field representation and functional saddle point}

The averaged resolvent of the normalized graph Laplacian (\ref{normLa}), to be used for constructing
the eigenvalue density as in (\ref{resolv}),
can be processed as follows
\beq
\langle\Tr(\pmb{\mathcal{L}}-z\mathbf{I})^{-1}\rangle=\langle\Tr[(1-z)\mathbf{I}-\mathbf{D}^{-1/2}\mathbf{A}\mathbf{D}^{-1/2}]^{-1}\rangle
=\langle\Tr\{\mathbf{D}[(1-z)\mathbf{D}-\mathbf{A}]^{-1}\}\rangle
\eeq
In view of the invariance of the \ER graph ensemble under vertex renumbering, and keeping in mind that the degree matrix $\mathbf{D}$ is diagonal with $D_{ii}\equiv\sum_{j\ne i}A_{ij}$, one can furthermore write
\beq\label{resolvN}
\begin{split}
\frac1N\langle\Tr(\pmb{\mathcal{L}}-z\mathbf{I})^{-1}\rangle
&=\langle D_{11}[(1-z)\mathbf{D}-\mathbf{A}]^{-1}_{11}\rangle=\langle({\textstyle\sum_{j>1}} A_{1j})\,[(1-z)\mathbf{D}-\mathbf{A}]^{-1}_{11}\rangle\\
&=(N-1)\langle A_{12}\,[(1-z)\mathbf{D}-\mathbf{A}]^{-1}_{11}\rangle
\end{split}
\eeq
We can then employ a Gaussian representation for the matrix inverse analogous to (\ref{Gausssuperv})
to obtain
\beq
[(1-z)\mathbf{D}-\mathbf{A}]^{-1}_{11}\propto \int d\pmb{\Psi}\,(i\phi_1^2)\, e^{i\sum_{k<l}A_{kl} [(1-z) (\Psi^\dagger_k\Psi_k+\Psi^\dagger_l\Psi_l)-2\Psi^\dagger_k \Psi_l]}.
\eeq
This expression has to be substituted into (\ref{resolvN}) and averaged over the \ER ensemble, which simply amounts to treating $A_{kl}$ with $k<l$ as independent random variables that equal 1 with probability $c/N$ and 0 with probability $1-c/N$. Note that the averaging over $A_{12}$ works slightly different from the other entries due to the explicit insertion of $A_{12}$ in (\ref{resolvN}). At the end of the day,
\begin{align*}
\langle\Tr(\pmb{\mathcal{L}}-z\mathbf{I})^{-1}\rangle\propto &
 \int d\pmb{\Psi}\,(i\phi_1^2)\, e^{i [(1-z) (\Psi^\dagger_1\Psi_1+\Psi^\dagger_2\Psi_2)-2\Psi^\dagger_1 \Psi_2]}\\
&\times\exp\Big\{\sum_{2\le k<l}\ln\Big[1+\frac{c}{N}\big(e^{i [(1-z) (\Psi^\dagger_k\Psi_k+\Psi^\dagger_l\Psi_l)-2\Psi^\dagger_k \Psi_l]}-1\big)\Big]\Big\}.
\end{align*}
At $N\to\infty$ this reduces to
\beq\label{GaussLn}
\begin{split}
\langle\Tr(\pmb{\mathcal{L}}-z\mathbf{I})^{-1}\rangle\propto &
 \int d\pmb{\Psi}\,(i\phi_1^2)\, e^{i [(1-z) (\Psi^\dagger_1\Psi_1+\Psi^\dagger_2\Psi_2)-2\Psi^\dagger_1 \Psi_2]}\\
&\times\exp\Big[-\frac{c}{2N}\sum_{kl} \big(1-e^{i[(1-z) (\Psi^\dagger_k\Psi_k+\Psi^\dagger_l\Psi_l)-2\Psi^\dagger_k \Psi_l]}\big)\Big].
\end{split}
\eeq
Note that, apart from the extra insertion that only affects the integrals over $\Psi_1$ and $\Psi_2$, but not the remaining $N-2$ integrals, the structure matches (\ref{Mav}) under the following identification:
one must set $z$ in (\ref{Mav}) to 0 and then assign 
\beq
C(\Psi,\Psi')=c\left(1-e^{i[(1-z) (\Psi^\dagger\Psi+\Psi'^\dagger\Psi')-2\Psi^\dagger \Psi']}\right).
\eeq
Thereafter, the machinery of the Fyodorov-Mirlin method gets to work, producing a saddle point equation
analogous to (\ref{sddlsymm}) in the form
\beq\label{sddlLnPsi}
g_*(\Psi^\dagger\Psi)=-\frac{ic}{\pi}\int d\Psi'\, \left(e^{i[(1-z) (\Psi^\dagger\Psi+\Psi'^\dagger\Psi')-2\Psi^\dagger \Psi']}-1\right)\,e^{ig_*(\Psi'^\dagger\Psi')}.
\eeq
We proceed with the established route of introducing polar coordinates and performing the Grassmannian integrals to rewrite this equation in the form
\beq\label{sddlLn}
g_*(\rho)=-ic\left(e^{i(1-z)\rho}-1-\rho\,e^{i(1-z)\rho}\int_0^\infty\hspace{-3mm}d\rho'\,\frac{J_1(2\sqrt{\rho\rho'})}{\sqrt{\rho\rho'}}\,e^{ig_*(\rho')+i(1-z)\rho'}\right).
\eeq
This equation is structurally similar to the corresponding equation (\ref{sddlLER}) for ordinary graph Laplacians, though the explicit shape of its solutions is rather different as we shall see below. Equation (\ref{sddlLn}) does not appear immediately tractable, and developing a theory of its solutions at general $c$ is an intriguing mathematical problem that we leave for future endeavors. Nonetheless, one can effectively
approximate the solutions of (\ref{sddlLn}) at large $c$, as we shall see below.

Because the integral representation (\ref{GaussLn}) contains extra insertions in the integrand compared to our general discussion around (\ref{Gausssuperv}), the saddle point estimate (\ref{resolvsddl}) will get modified. One can retrace the steps of the Fyodorov-Mirlin method to obtain, instead of (\ref{resolvintPsi}),
\beq
\langle\Tr(\pmb{\mathcal{L}}-z\mathbf{I})^{-1}\rangle\propto 
 \int d\Psi_1d\Psi_2\,(i\phi_1^2)\, e^{i [(1-z) (\Psi^\dagger_1\Psi_1+\Psi^\dagger_2\Psi_2)-2\Psi^\dagger_1 \Psi_2]}e^{ig_*(\Psi_1^\dagger\Psi_1)+ig_*(\Psi_2^\dagger\Psi_2)}.
\eeq
From (\ref{sddlLnPsi}),
$$
\int d\Psi_2\,e^{i[(1-z) (\Psi_1^\dagger\Psi_1+\Psi_2^\dagger\Psi_2)-2\Psi_1^\dagger \Psi_2]}\,e^{ig_*(\Psi_2^\dagger\Psi_2)}=\pi\left(1+\frac{i g_*(\Psi_1^\dagger\Psi_1)}{c}\right),
$$
so that
\beq
\langle\Tr(\pmb{\mathcal{L}}-z\mathbf{I})^{-1}\rangle\propto 
 \int d\Psi_1\,(i\phi_1^2)\left(1+\frac{i g_*(\Psi_1^\dagger\Psi_1)}{c}\right)e^{ig_*(\Psi_1^\dagger\Psi_1)}.
\eeq
Converting to the polar coordinates (\ref{polar}) and performing the integrals, we get
\beq
\langle\Tr(\pmb{\mathcal{L}}-z\mathbf{I})^{-1}\rangle\propto 
i \int_0^\infty\hspace{-3mm} d\rho\left(1+\frac{i g_*(\rho)}{c}\right)e^{ig_*(\rho)},
\eeq
and hence the eigenvalue density of the normalized Laplacian is given by
\beq\label{pLint}
p_{\mathcal{L}}(\lambda)\propto\Re\left[\int_0^\infty\hspace{-3mm} d\rho\left(1+\frac{i g_*(\rho)}{c}\right)e^{ig_*(\rho)}\right]\Bigg|_{z=\lambda}.
\eeq

%%%%%%%%%%%%

\subsection{Large $c$ asymptotics}

We have found the following scaling particularly useful in handling the large $c$ limit of equation (\ref{sddlLn}):
\beq\label{scalingLn}
z=1+\frac{\tilde z\sqrt{c}}{c+1},\qquad \rho=\frac{\tilde{\rho}}{\sqrt{c}},\qquad g_*(\rho(\tilde\rho),z(\tilde z))=\tilde g(\tilde\rho,\tilde z)-\tilde z\tilde\rho\frac{c}{c+1}.
\eeq
In the new variables, and dropping all tildes from now on,
\beq\label{sddlLnscaled}
g(\rho)=-ic\left(e^{-iz\rho/(c+1)}-1+\frac{iz\rho}{c+1}\right)+i\rho\,e^{-iz\rho/(c+1)}\int_0^\infty\hspace{-3mm}d\rho'\,\frac{J_1(2\sqrt{\rho\rho'/c})}{\sqrt{\rho\rho'/c}}\,e^{ig(\rho')-iz\rho'}.
\eeq
The large $c$ limit of this equation agrees exactly with (\ref{gstWig}), which gives rise to the Wigner semicircle. Thus, one expects a shifted and scaled Wigner semicircle at large $c$,\footnote{Emergence of the Wigner semicircle in this limit has been analyzed from a mathematical perspective in \cite{normWig}.} and the corrections, naively, are organized in powers of $1/c$, as opposed to the powers of $1/\sqrt{c}$ in the case of the ordinary Laplacian.

There is a subtlety, however. The Wigner semicircle has sharp edges, and if one attempts to build solutions of (\ref{sddlLnscaled}) as a naive
perturbative expansion in powers of $1/c$, one immediately runs into singularities at the edges. A simple prototype of this issue is seen if attempting to expand $\sqrt{1+1/c-x^2}$ as a power series in $1/c$. Spurious singularities emerge at $x=\pm 1$.

We will therefore have to be more ingenious in constructing an effective approximation to the solutions of
(\ref{sddlLnscaled}) at large $c$. Expanding (\ref{sddlLnscaled}) up to terms of order $1/c$, with $J_1(2y)=y-y^3/2+\cdots$, we get
$$
g(\rho)=\rho^2\left(\frac{icz^2}{2(c+1)^2}+\frac{z}{c+1}\int_0^\infty\hspace{-3mm}d\rho'e^{ig(\rho')-iz\rho'}-\frac{i}{2c}\int_0^\infty\hspace{-3mm}d\rho'\rho'e^{ig(\rho')-iz\rho'}\right)
+i\rho\int_0^\infty\hspace{-3mm}d\rho'e^{ig(\rho')-iz\rho'}.
$$
It is natural to look for a solution in the form
\beq\label{guv}
g(\rho,z)=-u(z)\rho+v(z)\rho^2,
\eeq
which yields (neglecting the difference between $1/(c+1)$ and $1/c$ in $v$)
\beq
\begin{split}
&v=\frac{1}{c}\left(\frac{iz^2}{2}+{z}\int_0^\infty\hspace{-3mm}d\rho\,e^{-i(u+z)\rho+iv\rho^2}-\frac{i}{2}\int_0^\infty\hspace{-3mm}d\rho\,\rho\,e^{-i(u+z)\rho+iv\rho^2}\right),\\ 
&u=-i\int_0^\infty\hspace{-3mm}d\rho\,e^{-i(u+z)\rho+iv\rho^2}.
\end{split}
\eeq
Since $v$ is of order $1/c$, one may attempt to approximate $e^{i v\rho^2}=1$ in the first line and $e^{i v\rho^2}=1+iv\rho^2$ in the second line, which leads to
\beq\label{vsol}
v=\frac{i}{c}\left(\frac{z^2}{2}-\frac{z}{u+z}+\frac{1}{2(u+z)^2}\right),\qquad
u^2+uz+1=\frac{2iv}{(u+z)^2}.
\eeq
If $v$ is neglected altogether in the second equation, the solution is $u(z)=W(z)$ given by (\ref{W}), in accord with convergence to the Wigner semicircle. We can use this estimate for $u$ to compute an estimate for $v$ from the first equation, and then substitute it in the second equation, obtaining
\beq
u^2+uz+1=-\frac{W^2}{c}(z^2+2zW+W^2),
\eeq
where, once again, we have approximated $u$ by $W$ in terms suppressed by $1/c$ and used $1/(W+z)=-W$. Using the equation for $W$ further, the right-hand side is simplified as
\beq\label{ueq}
u^2+uz+1=-\frac{1}{c}.
\eeq
This is solved by
\beq\label{usol}
u=\frac12\left(-z-i\sqrt{\frac{4(c+1)}{c}-z^2}\right).
\eeq
Unlike the naive perturbative expansion of $g$ in powers of $1/c$, this expression does not blow up near $z=\pm 1$. (And indeed, re-expanding it as a power series in $1/c$ would have immediately re-introduced the singularities, and should thus be avoided.)

Finally, from (\ref{pLint}) and (\ref{guv}), keeping in mind the redefinitions (\ref{scalingLn}), approximating $e^{i v\rho^2}=1+iv\rho^2$ and neglecting $v/c$,
\beq
p_{\mathcal{L}}(\lambda)\Big|_{\la=1+{\tilde z\sqrt{c}}/(c+1)}
\propto\Im\left[\frac{1-1/c}{u(z)+z c/(c+1)}-\frac{2iv(z)}{[u(z)+z c/(c+1)]^3}\right]
\eeq
The second term is of order $1/c$ so we can ignore the difference between $c/(c+1)$ and 1 in the denominator and then use (\ref{vsol}) and (\ref{ueq}) to write 
\beq\label{pLintfnl}
p_{\mathcal{L}}(\lambda)\Big|_{\la=1+{\tilde z\sqrt{c}}/(c+1)}
\propto-\Im\left[u(z)+\frac{u(z)^3}{c}\right].
\eeq
where $u$ is given by (\ref{usol}).

\begin{figure}[t]
\centering
\begin{subfigure}{0.45\textwidth}
\hspace{-1.5cm}\includegraphics[width = 1.2\linewidth]{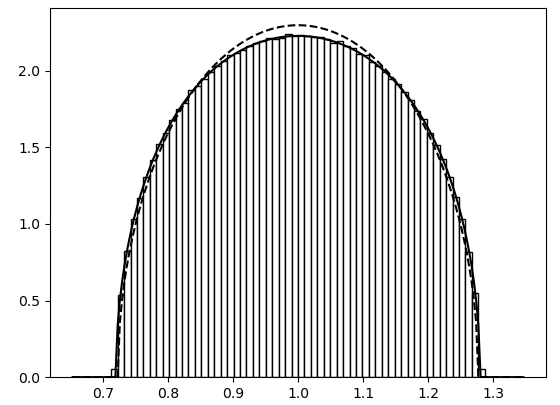}\vspace{-4mm}
\begin{picture}(0,0)
\put(-15,168){$p_{\mathcal{L}}$}
\put(190,19){$\la$}
\put(167,168){$c=50$}
\end{picture}
\caption{\rule{0.7cm}{0cm}}
\end{subfigure}
\begin{subfigure}{0.45\textwidth}
\hspace{-2mm}\includegraphics[width = 1.2\linewidth]{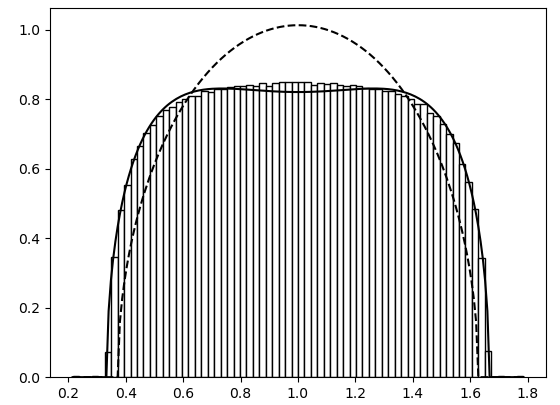}\vspace{-4mm}
\begin{picture}(0,0)
\put(22,168){$p_{\mathcal{L}}$}
\put(230,19){$\la$}
\put(210,168){$c=8$}
\end{picture}
\caption{\rule{-2cm}{0cm}}
\end{subfigure}
\caption{Theoretical prediction (\ref{pLint}) plotted as solid lines compared with the eigenvalue density of the normalized Laplacian
(\ref{normLa}) of an \ER random graph with $N=20000$ vertices and mean degree (a) $c=50$ and (b) $c=8$. The dashed lines represent the shifted and scaled Wigner semicircle that corresponds to discarding the $1/c$ corrections in (\ref{usol}) and (\ref{pLint}).}
\label{figLn}
\end{figure}
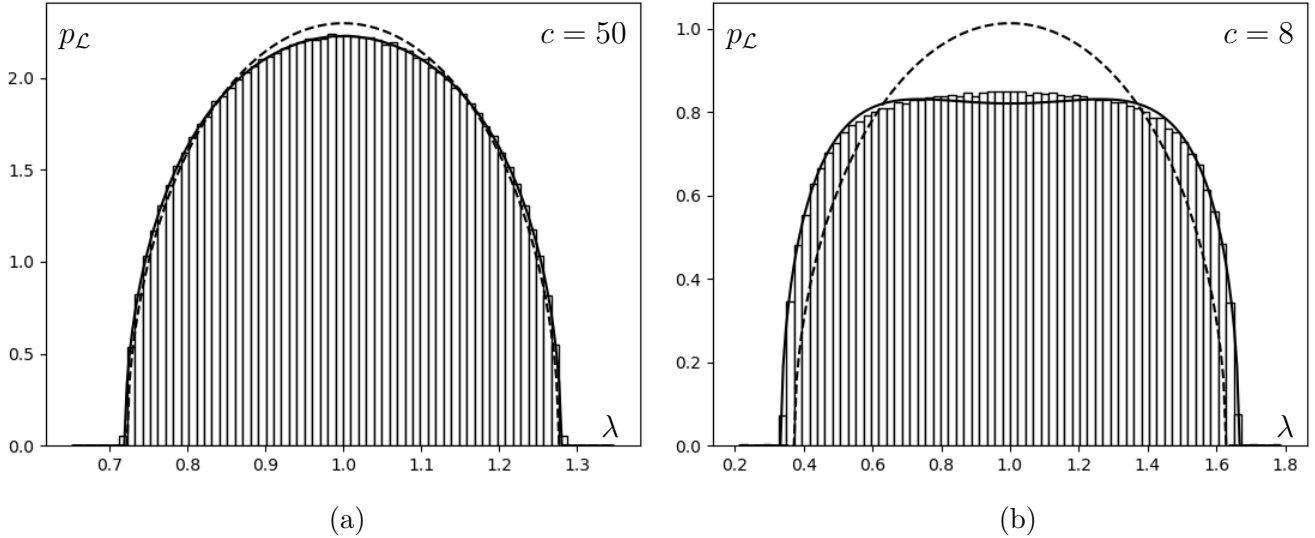
When $c$ is large, the second term in the above formula is negligible, while $u$ approaches $W$ so that the distribution is a scaled and shifted version of the Wigner semicircle (\ref{semicircle}). As an approximation to empirical distributions, shown in Fig.~\ref{figLn}, equation (\ref{pLintfnl}) remains adequate even for small values of $c$ a bit below 10, though deviations at the level of a few percent become visible at the lower end of this range. This is in accord with general expectations in terms of higher-order $1/c$ corrections.

%%%%%%%%%%%%%%%%%%%%%%%%%%%%%%%%%%%%%%%

\section{Outlook}

Using a supersymmetry-based resolvent representation and the Fyodorov-Mirlin method \cite{MF,FM},
we have developed an analytic theory for the eigenvalue distributions of random graph Laplacians.
The distributions can be extracted from solutions of explicit integral equations: (\ref{sddlLER}) for ordinary graph Laplacians and (\ref{sddlLn}) for normalized graph Laplacians. In the regime when the size of the graph tends to infinity, and the mean degree $c$ is fixed and large, asymptotic analysis in terms of $1/c$ can be performed, leading to explicit expressions for the eigenvalue density. As evident from Figs.~\ref{figL} and \ref{figLn}, these approximations correctly capture, at first subleading order, the empirically observed curves down to rather small values of $c$: $c\approx 20$ for ordinary Laplacians and $c\approx 10$ for normalized Laplacians. At larger values of $c$, our asymptotic estimates quickly become essentially exact. When $c$ tends to infinity, the distributions converge to shifted and scaled universal curves: the Wigner semicircle for the normalized graph Laplacian, and the Fyodorov distribution (\ref{pzrs}) for the ordinary graph Laplacian.

We mention a relation between our considerations and the function $\mathcal{F}(c)$ studied in mathematical literature \cite{span1,span2} and defined as
$\mathcal{F}(c)=\langle\ln(\tau(\mathbf{A}))\rangle/N$,
where $\tau(\mathbf{A})$ is the number of spanning trees of the graph given by the adjacency matrix $\mathbf{A}$. Since $\tau$ is expressed through the product of nonzero eigenvalues of the ordinary graph Laplacian (\ref{ordLa}) by the matrix tree theorem, one should have a relation of the form $\mathcal{F}(c)=\int d\lambda \ln(\la)\,p_L(\la)$ with $p_L$ given by (\ref{pL}), though extra care may have to be taken when handling the disconnected components of graphs at smaller values of $c$. In this way, our analytic results provide an alternative route to studying $\mathcal{F}(c)$.

Our derivations have definitely been executed at a physicist's level of rigor, with wishful assumptions
in relation to convergence of oscillatory integrals, contour deformations implicit in the saddle point method, justification of functional integration, identification of the dominant saddle points, etc. Our objective has been to use the established lore and plausible guesses to `craft' useful analytic formulas that compare favorably with the corresponding numerics. As far as this goal is concerned, Figs.~\ref{figL} and \ref{figLn} demonstrate that we have succeeded. We hope this initial step will pave the way for future rigorous mathematical work.

At a more practical level, getting some better mathematical understanding of the properties of the integral saddle point equations (\ref{sddlLER}) and (\ref{sddlLn}) would be very much in order. We have demonstrated
how to handle these equations effectively using asymptotic methods at large $c$. At the same time,
the equations contain a wealth of information beyond these asymptotic estimates. At our present level
of understanding, this information is not directly accessible, and even handling these equations numerically
is challenging due to the presence of oscillatory integrals. We hope the situation will improve in the future.
(Note that some related equations have been successfully treated numerically in \cite{gel}.)

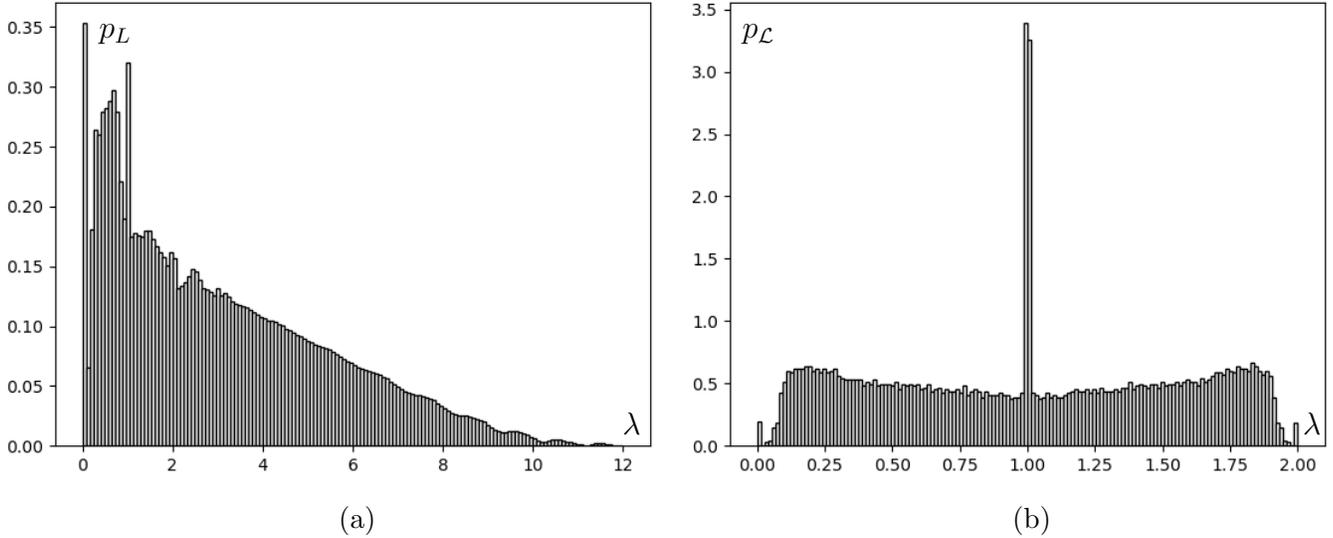
\begin{figure}[t]
\centering
\begin{subfigure}{0.45\textwidth}
\hspace{-1.3cm}\includegraphics[width = 1.2\linewidth]{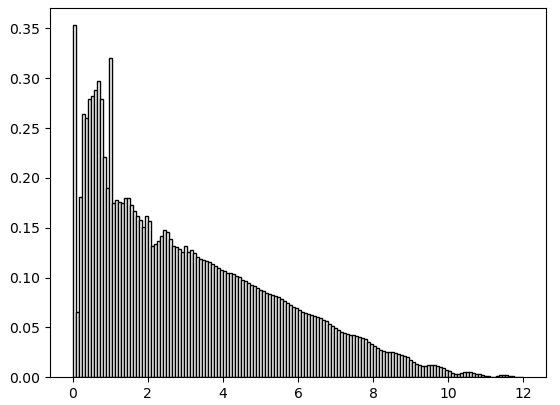}\vspace{-4mm}
\begin{picture}(0,0)
\put(2,170){$p_L$}
\put(200,20){$\la$}
\end{picture}
\caption{\rule{0.3cm}{0cm}}
\end{subfigure}
\begin{subfigure}{0.45\textwidth}
\hspace{1mm}\includegraphics[width = 1.2\linewidth]{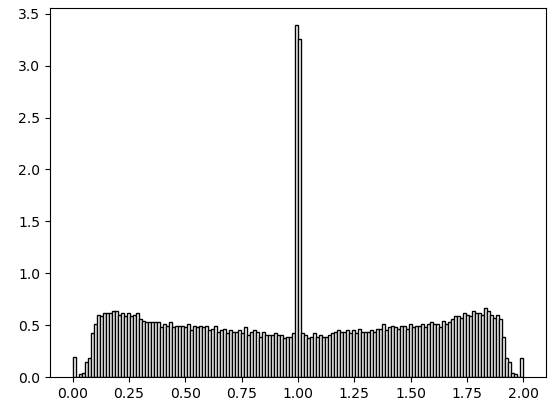}\vspace{-4mm}
\begin{picture}(0,0)
\put(30,170){$p_{\mathcal{L}}$}
\put(242,20){$\la$}
\end{picture}
\caption{\rule{-2.5cm}{0cm}}
\end{subfigure}
\caption{Empirical eigenvalue histograms for \ER random graphs with $c=3$, $N=5000$, averaged over 1000 instances: (a) ordinary Laplacian, (b) normalized Laplacian.}
\label{figc3}
\end{figure}
At smaller values of $c$, the Laplacian eigenvalue distributions attain more ornate shapes no longer captured by our asymptotic large $c$ analysis. In Fig.~\ref{figc3}, we provide the corresponding numerical histograms computed at $c=3$. As noted in \cite{mckay}, the curves for normalized Laplacians are very close to the McKay distribution for random $c$-regular graphs. 
This is intuitive, since, at large $c$, the degree variance in \ER graphs becomes small in comparison with $c$.
However, this relation with the McKay distribution does not appear under any explicit analytic control.
Finally, we mention that \cite{kuehn} discusses a very different approach to constructing eigenvalue distributions of sparse matrices and briefly applies it to the case of ordinary graph Laplacians. Considerations of \cite{kuehn} do not directly provide analytic expressions for the eigenvalue distributions (it may be possible in that framework to develop such analytics at large $c$), but rather
design a sampling scheme in the spirit of population dynamics that effectively reproduces the distribution
curve, in particular, at small $c$. It would be very interesting to unravel the relation between this approach
and the integral equations we have derived in this paper.\vspace{3mm}

\noindent {\bf Note added:} During the journal review of this article, a referee pointed out to us the earlier work in \cite{Khetal}, as well as the subsequent considerations in \cite{BL}. In \cite{Khetal}, an integral equation apparently equivalent to our (\ref{sddlLgen}) is derived for ordinary Laplacians of weighted graphs. The derivation, however, relies on intensive use of the method of moments and elaborate combinatorial counting (the considerations of \cite{Khetal} have been further strengthened and streamlined in \cite{BL}). Our single-page derivation in section~\ref{sddlLap}, based on the Fyodorov-Mirlin method, provides a nice complementary perspective. We remark further that the setting of 
section~\ref{sddlLap} is well-adapted to a number of relevant extensions, such as the analysis of eigenvalue density correlators, as in \cite{MF}.
In addition to deriving the integral saddle point equation in an economical way, we provided, in section~\ref{largecLap}, its asymptotic analysis, leading to an effective procedure to estimate the numerical values of its solutions. Finally, our analysis of normalized graph Laplacians
in section~\ref{secnorm} likewise reaches beyond the scope of \cite{Khetal}.

%%%%%%%%%%%%%%%%%%%%%%%%%%%%%%%%%%%%%%%

\section*{Acknowledgments}

We thank Peter Forrester for bringing \cite{zrs} to our attention; Pragya Shukla for correspondence; Eytan Katzav for encouragement, comments and pointing out the earlier work in \cite{kuehn}; Thip Chotibut for discussions and for drawing our attention to the McKay distribution and the way it is mentioned in \cite{mckay}. PA is funded by the SMART-UP scholarship program of Universit\'e Paris Cit\'e and has also been supported during the earlier stages of this work by Thailand Science Research and Innovation Fund Chulalongkorn University (IND66230005). OE is supported by Thailand NSRF via PMU-B (grant numbers B01F650006 and B05F650021). 

\appendix

\section*{Appendix: Python codes}

We provide below Python scripts that can be used to reproduce our main results displayed in Figs.~\ref{figL} and \ref{figLn}. While completely elementary, these scripts may be useful for curious readers, and they also serve as a formal justification for our empirical claims.

\subsection*{Ordinary graph Laplacian:}

\begin{verbatim}
import numpy as np
from scipy.integrate import odeint
from scipy.integrate import quad
from scipy.optimize import fsolve
import matplotlib.pyplot as plt

c=30.0
N=10000
p=c/(N-1)
sqrtc=c**0.5

def eq(f):
 integ,err=quad(lambda x: np.exp(-f*x-x**2/2),0,np.inf)
 return f-integ

f0=[fsolve(eq,0.1),0]

def fprime(f,z):
 fc=f[0]+1j*f[1]
 fpc=1j/(2+1j*z*fc-fc**2)
 return [np.real(fpc),np.imag(fpc)]

zhalf=np.linspace(0, 4, 101)
sol=odeint(fprime, f0, zhalf)
z=np.linspace(-4,4,201)
zscale=[c+x*sqrtc for x in z]
f=np.array([1+1j for i in range(201)])
for i in range(100):
 f[i]=sol[100-i,0]-1j*sol[100-i,1]
for i in range(100,201):
 f[i]=sol[i-100,0]+1j*sol[i-100,1]

fcorr=[f[i]-1j*(5*f[i]**4-11j*z[i]*f[i]**3-6*z[i]**2*f[i]**2-2*f[i]**2
       -6*z[i]**2-3j*z[i]*f[i]+2)/(6*sqrtc*(2+1j*z[i]*f[i]-f[i]**2)) 
       for i in range(201)]

freal=np.real(fcorr)
nrm=sum(freal)*sqrtc/25
frealnorm=freal/nrm

A=np.zeros((N,N))
for i in range(N-1):
 for j in range(i+1,N):
  if np.random.rand()<p:
   A[i,j]=-1
   A[j,i]=-1

D=np.sum(A,axis=0)
for i in range(N):
 A[i,i]=-D[i]

eig=np.linalg.eigvalsh(A)

plt.hist(eig, bins=70, range=(c-4*sqrtc,c+4*sqrtc),
         density=True, color='w', edgecolor='k')
plt.plot(zscale, frealnorm, color='k')
plt.show()
\end{verbatim}

\subsection*{Normalized graph Laplacian:}

\begin{verbatim}
import numpy as np
import matplotlib.pyplot as plt

c=30.0
N=10000
prob=c/(N-1)
sqrtc=c**0.5

z=np.linspace(-2.5,2.5,201)
zscale=[1+i*sqrtc/(c+1) for i in z]

u=[-(x+1j*(4*(1+1/c)+1e-10j*np.sign(x)-x**2)**0.5)/2 for x in z]
p=[-np.imag(u[i]+u[i]**3/c) for i in range(len(z))]
nrm=sum(p)*sqrtc/(c+1)*5/200
pnrm=p/nrm

A=np.zeros((N,N))
for i in range(N-1):
 for j in range(i+1,N):
  if np.random.rand()<prob:
   A[i,j]=1
   A[j,i]=1

D=np.sum(A,axis=0)
L=np.zeros((N,N))
for i in range(N):
 L[i,i]=1
 if D[i]>0:
  for j in range(i+1,N):
   if D[j]>0:
    L[i,j]=-A[i,j]/(D[i]*D[j])**0.5
    L[j,i]=L[i,j]

eig=np.linalg.eigvalsh(L)

plt.hist(eig, bins=70, range=(1-2.5*sqrtc/(c+1),1+2.5*sqrtc/(c+1)),
         density=True, color='w', edgecolor='k')
plt.plot(zscale, pnrm, color='k')
plt.show()
\end{verbatim}

%%%%%%%%%%%%%%%%%%%%%%%

\end{document}